\documentclass[fleqn,a4paper,11pt]{llncs}
\if@twoside
\else
\fi
\title{A Finite Equational Base for \CCS{} with Left Merge and
  Communication Merge%
   \thanks{The first and third author were partly supported by the
     project ``The Equational Logic of Parallel Processes''
     (nr.~060013021) of The Icelandic Research Fund.}
   \thanks{This is the full version of an extended abstract that has
     appeared as \cite{AFIL06}.}
}

\author{Luca Aceto\inst{1,4}
  \and
Wan Fokkink\inst{2,5}
\and
Anna Ingolfsdottir\inst{1,4}
\and
Bas Luttik\inst{3,5}
 }

\institute{%
  Department of Computer Science, Reykjav\'{\i}k University,
  Iceland%
  , \email{luca@ru.is,annai@ru.is}
\and
  Department of Computer Science,
  Vrije Universiteit Amsterdam,
  The Netherlands%
  , \email{wanf@cs.vu.nl}
\and
  Department of Mathematics and Computer Science,
  Technische Universiteit Eindhoven,\\
  The Netherlands%
 , \email{s.p.luttik@tue.nl}
\and
  {\bf BRICS},
  Department of Computer Science,
  Aalborg University,
  Denmark%
\and
  Department of Software Engineering,
  CWI,
  The Netherlands
   }
  \usepackage{amsmath}
  \usepackage{amssymb}
  \newcommand{\CCS}{\ensuremath{\mathsf{CCS}}}
  \newcommand{\ACP}[1][]{\ensuremath{\mathsf{ACP}_{#1}}}
  \newcommand{\VAR}[1][V]{\ensuremath{\mathcal{#1}}}
  \newcommand{\var}[1][x]{\ensuremath{\mathalpha{#1}}}
   \newcommand{\varx}{\var[x]}
   \newcommand{\vary}{\var[y]}
   \newcommand{\varz}{\var[z]}
  \newcommand{\Act}[1][A]{\ensuremath{\mathcal{#1}}}
  \newcommand{\Acts}[1][A]{\ensuremath{\mathcal{#1}_{\silent}}}
  \newcommand{\act}[1][\alpha]{\ensuremath{\mathalpha{#1}}}
   \newcommand{\acta}{\act[a]}
   \newcommand{\actb}{\act[b]}
  \let\co\overline
  \newcommand{\coact}[1][a]{\ensuremath{\mathalpha{\co{#1}}}}
   \newcommand{\coacta}{\coact[a]}
   \newcommand{\coactb}{\coact[b]}
   
  \newcommand{\nil}{\ensuremath{\mathalpha{\mathbf{0}}}}
  \newcommand{\silent}[1][\tau]{\ensuremath{\mathalpha{#1}}}
  \newcommand{\pref}[1]{\ensuremath{\mathalpha{{#1}.}}}
  
  \newcommand{\acmp}{\ensuremath{\mathbin{+}}}

  \newcommand{\pcmpsym}[1][]{\ensuremath{\mathalpha{\|_{#1}}}}
  \newcommand{\pcmp}[1][]{\ensuremath{\mathbin{\pcmpsym[#1]}}}
  \newcommand{\mergesym}[1][]{\ensuremath{\mathalpha{\|_{#1}}}}
  \newlength{\merged}
  \newlength{\mergeth}
  \newcommand{\lmergesym}[1][]{\ensuremath{\mathalpha{%
    \settoheight{\mergeth}{\mergesym}%
    \settodepth{\merged}{\mergesym}%
    \addtolength{\mergeth}{\merged}%
    \hbox{\rule[-\merged]{0pt}{\mergeth}%
          \setlength{\unitlength}{1ex}%
            \begin{picture}(1.94,2)%
            \thinlines%
            \put(.22,-.58){\line(1,0){1.5}}%
            \put(0,0){\mergesym}%
          \end{picture}%
  }_{#1}}}}
  \newcommand{\lmerge}[1][]{\ensuremath{\mathbin{\lmergesym[#1]}}}
  \newcommand{\cmergesym}[1][]{\ensuremath{|_{#1}}}
  \newcommand{\cmerge}[1][]{\ensuremath{\mathbin{\cmergesym[#1]}}}
  \newcommand{\hmerge}{\mathrel{|^{\negmedspace\scriptstyle /}}}
  \newcommand{\PTERMS}[1][P]{\ensuremath{\mathcal{#1}}}
  \newcommand{\pterm}[1][P]{\ensuremath{\mathalpha{#1}}}
   \newcommand{\ptermP}{\pterm[P]}
   \newcommand{\ptermQ}{\pterm[Q]}
   \newcommand{\ptermR}{\pterm[R]}
   \newcommand{\ptermS}{\pterm[S]}
   \newcommand{\ptermN}{\pterm[N]}
  \newcommand{\CPTERMS}[1][P]{\ensuremath{\mathcal{#1}_0}}
  \newcommand{\cpterm}[1][P]{\ensuremath{\mathalpha{#1}}}
   \newcommand{\cptermP}{\cpterm[P]}
   \newcommand{\cptermQ}{\cpterm[Q]}
   
     \newdimen\boxwdplusemdimen
    \def\arrow#1{{
      \boxwdplusemdimen=1em%
      \setbox0=\hbox{$\scriptstyle#1$}%
      \advance\boxwdplusemdimen by \wd0\relax%
      \ifdim\boxwdplusemdimen<16.11119pt%
        \boxwdplusemdimen=16.11119pt%
      \fi%
      \buildrel{#1}\over%
        {\setbox1=\hbox to \boxwdplusemdimen{\rightarrowfill}%
      \ht1=0.3em\relax\box1}%
    }}
  \newcommand{\step}[1]{\ensuremath{\mathbin{\arrow{#1}}}}
  \newcommand{\defined}[1]{\ensuremath{\mathalpha{{#1}\mathalpha{\downarrow}}}}
  \newcommand{\undefined}[1]{\ensuremath{\mathalpha{{#1}\mathalpha{\uparrow}}}}  \newcommand{\cfun}[1][\gamma]{\ensuremath{\mathalpha{#1}}}
    \newcommand{\fmcfun}{\ensuremath{\cfun[f]}}
    \newcommand{\ccscfun}{\ensuremath{\cfun[h]}}
  \newcommand{\brel}[1][R]{\ensuremath{\mathrel{\mathcal{R}}}}
  \newcommand{\bisimilar}[1][]{%
    \setbox0=\hbox{\kern-.1ex{$\leftrightarrow$}\kern-.1ex}
    \setbox1=\vbox{\hbox{\raise .1ex \box0}\hrule}%
    \ensuremath{\mathrel{\hbox{\kern.1ex\box1\kern.1ex}_{#1}}}
  }
  \newcommand{\bisimilargen}{\bisimilar[{\cfun}]}
  \newcommand{\eqclass}[1]{\ensuremath{[#1]}}
  \newcommand{\facalg}[2]{\ensuremath{{#1}{/}{#2}}} 
  \newcommand{\nila}{\nil}
  \newcommand{\prefa}[1]{\pref{#1}}
  \newcommand{\acmpa}{\acmp}
  \newcommand{\pcmpa}{\pcmp}
  \newcommand{\lmergea}{\lmerge}
  \newcommand{\cmergea}{\cmerge}
  \newcommand{\Proc}[1][\cfun]{\ensuremath{\mathbf{P}_{#1}}}
  \newcommand{\ProcFM}{\ensuremath{\Proc[\fmcfun]}}
  \newcommand{\ProcCCS}{\ensuremath{\Proc[\ccscfun]}}
  \newcommand{\proc}[1][p]{\ensuremath{#1}}
   \newcommand{\procp}{\proc[p]}
   \newcommand{\procq}{\proc[q]}
   \newcommand{\procr}{\proc[r]}
  \newcommand{\val}[1][\nu]{\ensuremath{#1}}
  \newcommand{\valstar}{\ensuremath{\val[*]}}
  \newcommand{\eval}[2][\val]{\ensuremath{{[\![}#2{]\!]}_{#1}}}
  
  \newcommand{\height}[1]{\ensuremath{h(#1)}}
  \newcommand{\length}[1]{\ensuremath{\ell(#1)}}
  
  \newcommand{\depth}[1]{\ensuremath{|#1|}}
  
  \newcommand{\bdeg}[1]{\ensuremath{\textit{bdeg}(#1)}}
  \newcommand{\fwidth}[1]{\ensuremath{\textit{w}(#1)}}
  \newcommand{\FMWidth}{\ensuremath{W}}
  \newcommand{\fpvarnum}[1]{\ensuremath{\ulcorner\!{#1}\!\urcorner}}
  \newcommand{\fvalstar}{\ensuremath{\val[*]}}
  \newcommand{\fevalstar}[2][\valstar]{\ensuremath{{[\![}#2{]\!]}_{#1}}}
  \newcommand{\bisimilarfm}{\bisimilar[{\fmcfun}]}
  \newcommand{\hwidth}[1]{\ensuremath{\textit{w}(#1)}}
  \newcommand{\HSWidth}{\ensuremath{W}}
  \newcommand{\HSAct}{\ensuremath{\Act[A']}}
  \newcommand{\hpvarnum}[1]{\ensuremath{\ulcorner\!{#1}\!\urcorner}}
  \newcommand{\hvalstar}{\ensuremath{\val[*]}}
  \newcommand{\hevalstar}[2][\valstar]{\ensuremath{{[\![}#2{]\!]}_{#1}}}
  \newcommand{\bisimilarccs}{\bisimilar[{\ccscfun}]}
  \newcommand{\EqTh}[1]{\ensuremath{\textit{EqTh}(#1)}}
  \newcommand{\EqBase}[1][E]{\ensuremath{\mathcal{#1}}}
  \newcommand{\fequate}[1][]{\ensuremath{\mathrel{\approx_{#1}}}}
   \newcommand{\feqfm}{\fequate[{\fmax{}}]}
   \newcommand{\feqhs}{\fequate[{\hsax{}}]}
  \newcommand{\IFF}{\ensuremath{\text{iff}}}
  \newcommand{\stepnl}{\ensuremath{\mathbin{\rightarrow}}}
  \newcommand{\stepnla}{\stepnl}
  \newcommand{\nstepnl}{\ensuremath{\mathalpha{\not\rightarrow}}}
  \newcommand{\nstepnla}{\nstepnl}
  
  \newcommand{\stepsnl}{\ensuremath{\mathbin{\rightarrow^*}}}
  \newcommand{\stepsnla}{\stepsnl}
  \newcommand{\brancher}[1][n]{\ensuremath{\varphi_{#1}}}
  \newcommand{\NFTERMS}[1][]{\ensuremath{\mathcal{N}_{#1}}}
   \newcommand{\fNFTERMS}{\NFTERMS[{\fmax{}}]}
   \newcommand{\hNFTERMS}{\NFTERMS[{\hsax{}}]}
  \newcommand{\N}{\ensuremath{\omega}}

  \spnewtheorem{thm}{Theorem}{\bfseries}{\upshape}
  \spnewtheorem{dfn}[thm]{Definition}{\bfseries}{\upshape}
  \spnewtheorem{lma}[thm]{Lemma}{\bfseries}{\upshape}
  \spnewtheorem{ppn}[thm]{Proposition}{\bfseries}{\upshape}
  \spnewtheorem{cry}[thm]{Corollary}{\bfseries}{\upshape}
  \spnewtheorem{rmk}[thm]{Remark}{\bfseries}{\upshape}
  \spnewtheorem{exe}[thm]{Example}{\bfseries}{\upshape}
  \spnewtheorem{ntn}[thm]{Notation}{\bfseries}{\upshape}
  \renewenvironment{theorem}{\begin{thm}}{\end{thm}}
  \renewenvironment{lemma}{\begin{lma}}{\end{lma}}
  \renewenvironment{proposition}{\begin{ppn}}{\end{ppn}}
  \renewenvironment{corollary}{\begin{cry}}{\end{cry}}

  \renewenvironment{definition}{\begin{dfn}}{\end{dfn}}
\begin{document}

  \maketitle

  \begin{abstract}
    Using the left merge and communication merge from \ACP{}, we
    present an equational base (i.e., a ground-complete and
    $\omega$-complete set of valid equations) for the fragment of
    \CCS{} without recursion, restriction and relabelling. Our
    equational base is finite if the set of actions is finite.
  \end{abstract}

  \section{Introduction} \label{sec:introduction}
  
  One of the first detailed studies of the equational theory of a
  process algebra was performed by Hennessy and Milner \cite{HM85}.
  They considered the equational theory of the process algebra that
  arises from the recursion-free fragment of \CCS{} (see
  \cite{Mil89}), and presented a set of equational axioms that is
  complete in the sense that all valid \emph{closed} equations (i.e.,
  equations in which no variables occur) are derivable from it in
  equational logic \cite{Tay79}. For the elimination of parallel
  composition from closed terms, Hennessy and Milner proposed the
  well-known \emph{Expansion Law}, an axiom schema that generates
  infinitely many axioms. Thus, the question arose whether a finite
  complete set of axioms exists. With their axiom system \ACP{},
  Bergstra and Klop demonstrated in \cite{BK84} that it does exist if
  two auxiliary operators are used: the left merge and the
  communication merge. It was later proved by Moller \cite{Mol90} that
  without using at least one auxiliary operator a finite complete set
  of axioms does not exist.
  
  The aforementioned results pertain to the closed fragments of the
  equational theories discussed, i.e., to the subsets consisting of
  the closed valid equations only. Many valid equations such as, e.g.,
  the equation
     $(\varx\pcmp\vary)\pcmp\varz
        \fequate
      \varx\pcmp(\vary\pcmp\varz)$
  expressing that parallel composition is associative, are not
  derivable (by means of equational logic) from the axioms in
  \cite{BK84} or \cite{HM85}.  In this paper we shall not neglect the
  variables and contribute to the study of full equational theories of
  process algebras. We take the fragment of \CCS{} without recursion,
  restriction and relabelling, and consider the full equational theory
  of the process algebra that is obtained by taking the syntax modulo
  bisimilarity \cite{Par81}.  Our goal is then to present an
  \emph{equational base} (i.e., a set of valid equations from which
  every other valid equation can be derived) for it, which is finite
  if the set of actions is finite. Obviously, Moller's result about
  the unavoidability of the use of auxiliary operations in a finite
  complete axiomatisation of the closed fragment of the equational
  theory of \CCS{} a fortiori implies that auxiliary operations are
  needed to achieve our goal. So we add left merge and communication
  merge from the start.
  
  Moller \cite{Mol89} considers the equational theory of the same
  fragment of \CCS{}, except that his parallel operator implements
  pure interleaving instead of \CCS{}-communi\-cation and the
  communication merge is omitted. He presents a set of valid 
  axiom schemata and proves that it generates an equational base if
  the set of actions is infinite. Groote \cite{Gro90} does consider
  the fragment including communication merge, but, instead of the
  \CCS{}-communication mechanism, he assumes an uninterpreted
  communication function. His axiom schemata also generate an
  equational base provided that the set of actions is infinite. We
  improve on these results by considering the communication mechanism
  present in \CCS{}, and by proving that our axiom schemata generate
  an equational base also if the set of actions is finite.  Moreover,
  our axiom schemata generate a finite equational base if the set of
  actions is finite.

  Our equational base consists of axioms that are mostly well-known.
  For parallel composition ($\pcmp{}$), left merge ($\lmerge$) and
  communication merge ($\cmerge$) we adapt the axioms of \ACP{},
  adding from Bergstra and Tucker \cite{BT85} a selection of the
  axioms for \emph{standard concurrency} and the axiom 
    $(\varx\cmerge\vary)\cmerge\varz \fequate \nil$,
  which expresses that the communication mechanism is a form of
  \emph{handshaking communication}. 

  Our proof follows the classic two-step approach: first we identify a
  set of normal forms such that every process term has a provably
  equal normal form, and then we demonstrate that for distinct normal
  forms there is a distinguishing valuation that proves that they
  should not be equated.  (We refer to the survey \cite{AFIL05} for a
  discussion of proof techniques and an overview of results and open
  problems in the area. We remark in passing that one of our main
  results in this paper, viz.~Corollary~\ref{cor:hscomplete}, solves
  the open problem mentioned in~\cite[p.~362]{AFIL05}.)  Since both
  associating a normal form with a process term and determining a
  distinguishing valuation for two distinct normal forms are easily
  seen to be computable, as a corollary to our proof we get the
  decidability of the equational theory.  Another consequence of our
  result is that our equational base is complete for the set of valid
  closed equations as well as $\omega$-complete \cite{Hee86}.

  The positive result that we obtain in Corollary~\ref{cor:hscomplete}
  of this paper stands in contrast
  with the negative result that we have obtained in \cite{AFIL05b}. In
  that article we proved that there does not exist a finite equational
  base for \CCS{} if the auxiliary operation $\hmerge$ of Hennessy
  \cite{Hen88} is added instead of Bergstra and Klop's left merge and
  communication merge.  Furthermore, we conjecture that a finite
  equational base fails to exist if the unary action prefixes are
  replaced by binary sequential composition. (We refer to
  \cite{AFIL05} for an infinite family of valid equations that we
  believe cannot all be derivable from a single finite set of valid
  equations.)

  The paper is organised as follows. In Sect.~\ref{sec:preliminaries}
  we introduce a class of algebras of processes arising from a
  process calculus \`a la \CCS{}, present a set of equations that is
  valid in all of them, and establish a few general properties needed
  in the remainder of  the paper. Our class of process algebras is
  parametrised by a communication function. It is beneficial to
  proceed in this generality, because it allows us to first consider
  the simpler case of a process algebra with pure interleaving (i.e.,
  no communication at all) instead of \CCS{}-like parallel
  composition. In Sect.~\ref{sec:interleaving} we prove that an
  equational base for the process algebra with pure
  interleaving is obtained by simply adding the axiom
    $\varx\cmerge\vary\fequate\nil$
  to the set of equations introduced in Sect.~\ref{sec:preliminaries}.
  The proof in Sect.~\ref{sec:interleaving} extends nicely to a proof
  that, for the more complicated case of \CCS{}-communication, it is
  enough to replace
    $\varx\cmerge\vary\fequate\nil$
  by
    $\varx\cmerge(\vary\cmerge\varz)\fequate\nil$;
  this is discussed in Sect.~\ref{sec:ccsparallel}.

  \section{Algebras of Processes} \label{sec:preliminaries}

  We fix a set $\Act$ of \emph{actions}, and declare a special action
  $\silent$ that we assume is not in $\Act$.
  We denote by $\Acts$ the set $\Act\cup\{\silent\}$.
  Generally, we let $\acta$ and $\actb$ range over $\Act$ and
  $\act$ over $\Acts$.
  We also fix a countably infinite set $\VAR$ of \emph{variables}. The
  set $\PTERMS$ of \emph{process terms} is generated by the following
  grammar:
  \begin{equation*}
    \ptermP ::=\ \varx\ \mid\
                \nil\ \mid\
                \pref{\act}\ptermP\ \mid\
                \ptermP\acmp\ptermP\ \mid\
                \ptermP\lmerge\ptermP\ \mid\
                \ptermP\cmerge\ptermP\ \mid\
                \ptermP\pcmp\ptermP
  \enskip,
  \end{equation*}
  with $\varx\in\VAR$, and $\act\in\Acts$.
  We shall often simply write $\act$ instead of
  $\pref{\act}\nil$. Furthermore, to be able to omit some parentheses
  when writing terms, we adopt the convention that $\pref{\act}$ binds
  stronger and $\acmp$ binds weaker than all the other operations.

  \begin{table}
  \center\small
  \caption{The operational semantics.}
    \label{tab:tssccs}
  \rule{\textwidth}{.4pt}
  \begin{tabular}{c} \\
    $\dfrac{}{\pref{\act}\cptermP\step{\act}\cptermP}$
  \quad
      $\dfrac{\cptermP\step{\act}\cpterm[P']}
             {\cptermP\acmp\cptermQ\step{\act}\cpterm[P']}$
    \quad
      $\dfrac{\cptermQ\step{\act}\cpterm[Q']}
             {\cptermP\acmp\cptermQ\step{\act}\cpterm[Q']}$
\\ \\
    $\dfrac{\cptermP\step{\act}\cpterm[P']}
           {\cptermP\lmerge\cptermQ\step{\act}\cpterm[P']\pcmp\cptermQ}$
  \quad
    $\dfrac{\cptermP\step{\act}\cpterm[P']}
           {\cptermP\pcmp\cptermQ\step{\act}\cpterm[P']\pcmp\cptermQ}$
  \quad
    $\dfrac{\cptermQ\step{\act}\cpterm[Q']}
           {\cptermP\pcmp\cptermQ\step{\act}\cptermP\pcmp\cpterm[Q']}$
\\ \\
    $\dfrac{\cptermP\step{\acta}\cpterm[P'],\
            \cptermQ\step{\actb}\cpterm[Q'],\
            \defined{\cfun(\acta,\actb)}}
           {\cptermP\cmerge\cptermQ\step{\cfun(\acta,\actb)}\cpterm[P']\pcmp\cpterm[Q']}$
  \quad
    $\dfrac{\cptermP\step{\acta}\cpterm[P'],\
            \cptermQ\step{\actb}\cpterm[Q'],\
            \defined{\cfun(\acta,\actb)}}
           {\cptermP\pcmp\cptermQ\step{\cfun(\acta,\actb)}\cpterm[P']\pcmp\cpterm[Q']}$
\\ \\
    \end{tabular}
  \rule{\textwidth}{.4pt}
  \end{table}
  
  A process term is \emph{closed} if it does not contain variables; we
  denote the set of all closed process terms by $\CPTERMS$.
  We define on $\CPTERMS$ binary relations $\step{\act}$
  ($\act\in\Acts$) by means of the transition system specification in 
  Table~\ref{tab:tssccs}. The last two rules in Table~\ref{tab:tssccs}
  refer to a \emph{communication function} $\cfun$, i.e., a
  commutative and associative partial binary function
    $\cfun:\Act\times\Act\rightharpoonup\Acts$.
  We shall abbreviate the statement `$\cfun(\acta,\actb)$ is defined'
  by $\defined{\cfun(\acta,\actb)}$ and the statement
  `$\cfun(\acta,\actb)$ is undefined' by $\undefined{\cfun(\acta,\actb)}$.
  We shall in particular consider the following communication functions:
  \begin{enumerate}
  \item The \emph{trivial communication function} is the partial
    function
      $\fmcfun:\Act\times\Act\rightharpoonup\Acts$
    such that
      $\undefined{\fmcfun(\acta,\actb)}$
        for all $\acta,\actb\in\Act$.
  \item The \emph{\CCS{} communication function}
      $\ccscfun:\Act\times\Act\rightharpoonup\Acts$ 
    presupposes a bijection $\bar{.}$ on $\Act$ such that
    $\co{\co{\acta}}=\acta$ and $\co{\acta}\not=\acta$ for all
    $\acta\in\Act$, and is then defined by
    $\ccscfun(\acta,\actb)=\silent$ if $\co{\acta}=\actb$ and
    undefined otherwise.
  \end{enumerate}

  \begin{definition}
    A \emph{bisimulation} is a symmetric binary relation $\brel$ on
    $\CPTERMS$ such that $\cptermP\brel\cptermQ$ implies
    \begin{equation*}
    \text{if $\cptermP\step{\act}\cpterm[P']$, then there exists
      $\cpterm[Q']\in\CPTERMS$ such that $\cptermQ\step{\act}\cpterm[Q']$
      and $\cpterm[P']\brel\cpterm[Q']$.}
    \end{equation*}
    Closed process terms $\cptermP,\cptermQ\in\CPTERMS$ are said to be
    \emph{bisimilar} (notation: $\cptermP\bisimilargen\cptermQ$) if
    there exists a bisimulation $\brel$ such that
    $\cptermP\brel\cptermQ$.
  \end{definition}
  The relation $\bisimilargen$ is an equivalence relation on $\CPTERMS$;
  we denote the equivalence class containing $\cpterm$ by
  $\eqclass{\cpterm}$, i.e.,
  \begin{equation*}
    \eqclass{\cpterm}=\{\cptermQ\in\CPTERMS : \cptermP\bisimilargen\cptermQ\}
  \enskip.
  \end{equation*}
  The rules in Table~\ref{tab:tssccs} are all in de Simone's format
  \cite{Sim85} if $\ptermP$, $\pterm[P']$, $\ptermQ$ and $\pterm[Q']$
  are treated as variables ranging over closed process terms and
  the last two rules are treated as rule schemata generating a rule for
  all $\acta$, $\actb$ such that
  $\defined{\cfun(\acta,\actb)}$. Hence, $\bisimilargen$ has the
  substitution property for the syntactic constructs of our language
  of closed process terms, and therefore the constructs induce an
  algebraic structure on $\facalg{\CPTERMS}{\bisimilargen}$, with a
  constant $\nila$, unary operations $\prefa{\act}$ ($\act\in\Acts$)
  and four binary operations $\acmpa$, $\lmergea$, $\cmergea$ and
  $\pcmpa$ defined by
  \begin{alignat*}{2}
  & \nila = \eqclass{\nil}
  &&\qquad
    \eqclass{\cptermP}\lmergea\eqclass{\cptermQ}=
      \eqclass{\cptermP\lmerge\cptermQ}
  \\
  & \prefa{\act}\eqclass{\cptermP}=
      \eqclass{\pref{\act}\cptermP}
  &&\qquad
    \eqclass{\cptermP}\cmergea\eqclass{\cptermQ}=
      \eqclass{\cptermP\cmerge\cptermQ}
  \\
  & \eqclass{\cptermP}\acmpa\eqclass{\cptermQ}=
      \eqclass{\cptermP\acmp\cptermQ}
  &&\qquad
    \eqclass{\cptermP}\pcmpa\eqclass{\cptermQ}=
      \eqclass{\cptermP\pcmp\cptermQ}
  \enskip.
  \end{alignat*}%
  Henceforth, we denote by $\Proc$ (for $\cfun$ an arbitrary
  communication function) the algebra obtained by dividing
  out $\bisimilargen$ on $\CPTERMS$ with constant $\nila$ and operations
  $\prefa{\act}$ ($\act\in\Acts$), $\acmpa$, $\lmergea$, $\cmergea$,
  and $\pcmpa$ as defined above.
  The elements of $\Proc$ are called \emph{processes}, and will be
  ranged over by $\procp$, $\procq$ and $\procr$.

  \subsection{Equational Reasoning} \label{sec:eqlog}

\newcommand{\bccspax}[1]{\ensuremath{\mathrm{A#1}}}
 \newcommand{\accom}{\bccspax{1}}
 \newcommand{\acass}{\bccspax{2}}
 \newcommand{\acidp}{\bccspax{3}}
 \newcommand{\acnil}{\bccspax{4}}
\newcommand{\fmergelm}{\ensuremath{\mathrm{M}}}
\newcommand{\pcmpax}[1]{\ensuremath{\mathrm{P#1}}}
 \newcommand{\pcmplmcm}{\pcmpax{1}}
\newcommand{\lmergeax}[1]{\ensuremath{\mathrm{L#1}}}
 \newcommand{\nillm}{\lmergeax{1}}
 \newcommand{\preflm}{\lmergeax{2}}
 \newcommand{\aclm}{\lmergeax{3}}
 \newcommand{\lmlm}{\lmergeax{4}}
 \newcommand{\lmnil}{\lmergeax{5}}
\newcommand{\cmergeax}[1]{\ensuremath{\mathrm{C#1}}}
 \newcommand{\nilcm}{\cmergeax{1}}
 \newcommand{\prefcmd}{\cmergeax{2}}
 \newcommand{\prefcmu}{\cmergeax{3}}
 \newcommand{\accm}{\cmergeax{4}}
 \newcommand{\cmcom}{\cmergeax{5}}
 \newcommand{\cmass}{\cmergeax{6}}
 \newcommand{\cmlm}{\cmergeax{7}}
 \newcommand{\lmcmlm}{\cmergeax{8}}
\newcommand{\hsax}{\ensuremath{\mathrm{H}}}
\newcommand{\fmax}{\ensuremath{\mathrm{F}}}

  We can use the full language of process expressions to reason about
  the elements of $\Proc$.
  A \emph{valuation} is a mapping $\val:\VAR\rightarrow\Proc$; it
  induces an \emph{evaluation mapping}
  \begin{equation*}
    \eval{\_}:\PTERMS\rightarrow\Proc
  \end{equation*}
  inductively defined by
  \begin{alignat*}{2}
  & \eval{\varx} = \val(\varx)
  &&\qquad \eval{\ptermP\lmerge\ptermQ} = \eval{\ptermP}\lmergea\eval{\ptermQ} \\
  & \eval{\nil} = \nila
  &&\qquad \eval{\ptermP\cmerge\ptermQ} = \eval{\ptermP}\cmergea\eval{\ptermQ} \\
  & \eval{\pref{\act}\ptermP} = \prefa{\act}\eval{\ptermP}
  &&\qquad \eval{\ptermP\pcmp\ptermQ} =
  \eval{\ptermP}\pcmpa\eval{\ptermQ}
\\
  & \eval{\ptermP\acmp\ptermQ} = \eval{\ptermP}\acmp\eval{\ptermQ}.
  \end{alignat*}%
  A \emph{process equation} is a formula $\ptermP\fequate\ptermQ$
  with $\ptermP$ and $\ptermQ$ process terms; it is said to be
  \emph{valid} (in $\Proc$) if $\eval{\ptermP}=\eval{\ptermQ}$ for all
  $\val:\VAR\rightarrow\Proc$.
  If $\ptermP\fequate\ptermQ$ is valid in $\Proc$, then we shall also write
  $\ptermP\bisimilargen\ptermQ$. The \emph{equational theory} of
  the algebra $\Proc$ is the set of all valid process equations, i.e.,
  \begin{equation*}
    \EqTh{\Proc}=
      \{\ptermP\fequate\ptermQ:
          \text{$\eval{\ptermP}=\eval{\ptermQ}$
                  for all $\val:\VAR\rightarrow\Proc$}
      \}
  \enskip.   
  \end{equation*}
  The precise contents of the set $\EqTh{\Proc}$ depend to some
  extent on the choice of $\cfun$. For instance, the process equation
    $\varx\cmerge\vary\ \fequate\ \nil$
  is only valid in $\Proc$ if $\cfun$ is the trivial communication
  function \fmcfun{}; if $\cfun$ is the \CCS{} communication function
  \ccscfun{}, then $\Proc$ satisfies the weaker equation
    $\varx \cmerge (\vary \cmerge \varz)\ \fequate\ \nil$.

  \begin{table}
  \caption{Process equations valid in every \Proc{}.}
  \label{tab:axccs}
  \rule{\textwidth}{.4pt}
  \begin{center}
  \begin{tabular}{@{}ll@{}}
    $\begin{array}[t]{@{}l@{\quad}l@{\ \fequate\ }l@{}}
      \accom{}  & \varx \acmp \vary & \vary \acmp \varx
    \\
      \acass{}  & (\varx \acmp \vary) \acmp \varz
                    &
                  \varx \acmp (\vary \acmp \varz)
    \\
      \acidp{}  & \varx \acmp \varx
                    &
                  \varx
    \\
      \acnil{}  & \varx \acmp \nil & \varx
    \\ \\
      \nillm{}  & \nil \lmerge \varx & \nil
    \\
      \preflm{} & \pref{\act}\varx\lmerge\vary
                    &
                  \pref{\act}(\varx\pcmp\vary)
    \\
      \aclm{}   & (\varx\acmp\vary)\lmerge\varz
                    &
                  \varx \lmerge \varz \acmp \vary \lmerge \varz
    \\
      \lmlm{}   & (\varx \lmerge \vary)\lmerge \varz
                    &
                  \varx \lmerge (\vary \pcmp \varz)
    \\
      \lmnil{}  & \varx \lmerge \nil & \varx
    \end{array}$ &\qquad
    $\begin{array}[t]{@{}l@{\quad}l@{\ \fequate\ }l@{\quad}l@{}}
      \nilcm{}  & \nil \cmerge \varx & \nil
    \\
      \prefcmd{} & \pref{\acta}\varx\cmerge\pref{\actb}\vary
                    &
                  \pref{\cfun(\acta,\actb)}(\varx\pcmp\vary)
        & \text{if $\defined{\cfun(\acta,\actb)}$}
    \\
      \prefcmu{} & \pref{\acta}\varx\cmerge\pref{\actb}\vary
                    &
                  \nil
        & \text{if $\undefined{\cfun(\acta,\actb)}$}
    \\
      \accm{}   & (\varx\acmp\vary)\cmerge\varz
                    &
                  \varx \cmerge \varz \acmp \vary \cmerge \varz
    \\
      \cmcom{}  & \varx\cmerge\vary & \vary\cmerge\varx
    \\
      \cmass{}  & (\varx \cmerge \vary)\cmerge\varz
                    &
                  \varx\cmerge(\vary\cmerge\varz)
    \\
      \cmlm{}   & (\varx\lmerge\vary) \cmerge \varz
                    &
                  (\varx \cmerge \varz) \lmerge \vary
    \\ \\
      \pcmplmcm{} & \varx \pcmp \vary
                     &
                   \multicolumn{2}{l}{(\varx\lmerge\vary \acmp \vary\lmerge\varx)
                     \acmp \varx\cmerge\vary}
    \end{array}$ \\
  \end{tabular}
  \end{center}
  \rule{\textwidth}{.4pt}
  \end{table}

  Table~\ref{tab:axccs} lists process equations that are valid in
  $\Proc$ independently of the choice of $\cfun$. (The equations
  \preflm{}, \prefcmd{} and \prefcmu{} are actually axiom schemata;
  they generate an axiom for all $\act\in\Acts$ and
  $\acta,\actb\in\Act$. Note that if $\Act$ is finite, then these
  axiom schemata generate finitely many axioms.)
  Henceforth whenever we write an equation $\ptermP\fequate\ptermQ$,
  we mean that it is derivable from the axioms in
  Table~\ref{tab:axccs} by means of equational logic.
  It is well-known that the rules of equational logic preserve
  validity.
  We therefore obtain the following result.

  \begin{proposition} \label{prop:gensound}
    For all process terms $\ptermP$ and $\ptermQ$, if
    $\ptermP\fequate\ptermQ$, then $\ptermP\bisimilargen\ptermQ$.
  \end{proposition}

  In the following lemma we give an example of a valid equation that
  can be derived from Table~\ref{tab:axccs} using the rules of
  equational logic.
  \begin{lemma} \label{lem:lmcmlm}
    The following equation is derivable from the axioms in
    Table~\ref{tab:axccs}:
    \begin{equation*}
      \begin{array}[t]{@{}l@{\quad}l@{\ \fequate\ }l@{}}
        \lmcmlm & (\varx\lmerge\vary)\cmerge(\varz\lmerge\var[u])
                & (\varx\cmerge\varz)\lmerge(\vary\pcmp\var[u])
    \enskip.
      \end{array}
    \end{equation*}
  \end{lemma}%
  \begin{proof}
    The lemma is proved with the derivation:
    \begin{alignat*}{2}
      (\varx\lmerge\vary)\cmerge(\varz\lmerge\var[u])
    & \fequate
      (\varz\lmerge\var[u])\cmerge(\varx\lmerge\vary)
    &&\qquad\text{(by \cmcom{})}\\
    & \fequate
      (\varz\cmerge(\varx\lmerge\vary))\lmerge\var[u]
    &&\qquad\text{(by \cmlm{})}\\
    & \fequate
      ((\varx\lmerge\vary)\cmerge\varz)\lmerge\var[u]
    &&\qquad\text{(by \cmcom{})}\\
    & \fequate
      ((\varx\cmerge\varz)\lmerge\vary)\lmerge\var[u]
    &&\qquad\text{(by \cmlm{})}\\
    & \fequate
      (\varx\cmerge\varz)\lmerge(\vary\pcmp\var[u])
    &&\qquad\text{(by \lmlm{}).}
    \end{alignat*}
  \qed
  \end{proof}

  A set of valid process equations is an \emph{equational
    base} for $\Proc$ if all other valid process equations are
  derivable from it by means of equational logic.
  The purpose of this paper is to prove that if we add to the
  equations in Table~\ref{tab:axccs} the equation
  $\varx\cmerge\vary\fequate\nil$ we obtain an equational base for
  $\ProcFM$, and if, instead, we add
  $\varx\cmerge(\vary\cmerge\varz)\fequate\nil$ we obtain an
  equational base for $\ProcCCS$. Both these equational bases are
  finite, if the set of actions $\Act$ is finite.

  For the proofs of these results, we adopt the classic two-step
  approach \cite{AFIL05}: 
  \begin{enumerate}
  \item In the first step we identify a set of normal forms, and prove
    that every process term can be rewritten to a normal form by means
    of the axioms. 
  \item In the second step we prove that bisimilar normal forms are
    identical modulo applications of the axioms
    \accom{}--\acnil{}. This is done by associating with every pair of
    normal forms a so-called distinguishing valuation, i.e., a
    valuation that proves that the normal forms are not bisimilar
    unless they are provably equal modulo the axioms
    \accom{}--\acnil{}.
  \end{enumerate}
  Many of the proofs to follow will be by induction using of the
  following syntactic measure on process terms.

  \begin{definition} \label{def:height}
  Let $\ptermP$ be a process term.
  We define the \emph{height} of a process term $\ptermP$,
  denoted $\height{\ptermP}$, inductively as follows:
  \begin{equation*}
  \begin{array}[t]{@{}l@{}}
    \height{\nil} = 0
  \enskip, \\
    \height{\varx} = 1
  \enskip, \\
    \height{\pref{\act}\ptermP} = \height{\ptermP}+1
  \enskip, \\
   \height{\ptermP\acmp\ptermQ} = \max(\height{\ptermP},\height{\ptermQ})
  \enskip.
  \end{array} \qquad
  \begin{array}[t]{@{}l@{}}
    \height{\ptermP\lmerge\ptermQ} = \height{\ptermP}+\height{\ptermQ}
  \enskip, \\
   \height{\ptermP\cmerge\ptermQ} = \height{\ptermP}+\height{\ptermQ}
  \enskip, \\
    \height{\ptermP\pcmp\ptermQ} = \height{\ptermP}+\height{\ptermQ}
  \enskip,
  \end{array}
  \end{equation*}
  \end{definition}

  \begin{definition} \label{def:simple}
    We call a process term \emph{simple} if it is not $\nil$ and not
    an alternative composition.
  \end{definition}

  \begin{lemma} \label{lem:simplify}
    For every process term $\ptermP$ there exists a collection of
    simple process terms $\pterm[S_1],\dots,\pterm[S_n]$ ($n\geq 0$)
    such that $\height{\ptermP}\geq\height{\pterm[S_i]}$ for all
    $i=1,\dots,n$ and
    \begin{equation*}
       \ptermP \fequate
         \sum_{i=1}^n \pterm[S_i]
    \qquad\text{(by \accom{}, \acass{} and \acnil{}).}
    \end{equation*}
    We postulate that the summation of an empty collection of terms
    denotes $\nil$. The terms $\pterm[S_i]$ will be called
    \emph{syntactic summands} of $\ptermP$.
  \end{lemma}

  \subsection{General Properties of $\Proc{}$} \label{subsec:procprops}

  We collect some general properties of the algebras $\Proc$ that we
  shall need in the remainder of the paper.

  The binary transition relations $\step{\act}$ ($\act\in\Acts$) on
  \CPTERMS{}, which were used to associate an operational semantics
  with closed process terms, will play an important r\^ole in the
  remainder of the paper. They induce binary relations on $\Proc$,
  also denoted by $\step{\act}$, and defined as the least relations
  such that
    $\cptermP\step{\act}\cpterm[P']$
  implies
    $\eqclass{\cptermP}\step{\act}\eqclass{\cpterm[P']}$.
  Note that we then get, directly from the definition of bisimulation,
  that for all $\cptermP,\cpterm[P']\in\CPTERMS$:
  \begin{equation*}
    \eqclass{\cptermP}\step{\act}\eqclass{\cpterm[P']}\
  \IFF{}\
    \text{for all $\cptermQ\in\eqclass{\cptermP}$ there exists
      $\cpterm[Q']\in\eqclass{\cpterm[P']}$ such that
        $\cptermQ\step{\act}\cpterm[Q']$.}
  \end{equation*}

  \begin{proposition} \label{prop:steplifting}
    For all $\procp,\procq,\procr\in\Proc$:
    \begin{enumerate}
    \renewcommand{\theenumi}{\alph{enumi}}
    \renewcommand{\labelenumi}{(\theenumi)}
    \item \label{prop:steplifting:item:nil}
      $\procp=\nila$
         \IFF{}
      there do not exist $\proc[p']\in\Proc$ and $\act\in\Acts$
      such that
        $\procp\step{\act}\proc[p']$;
    \item \label{prop:steplifting:item:pref}
      $\pref{\act}\procp\step{\act[\beta]}\procr$
        \IFF{}
      $\act=\act[\beta]$ and $\procr=\procp$;
    \item \label{prop:steplifting:item:acmp}
      $\procp\acmpa\procq\step{\act}\procr$
        \IFF{}
      $\procp\step{\act}\procr$ or $\procq\step{\act}\procr$;
    \item \label{prop:steplifting:item:lmerge}
      $\procp\lmergea\procq\step{\act}\procr$
        \IFF{}
      there exists $\proc[p']\in\Proc$ such that
        $\procp\step{\act}\proc[p']$
          and
        $\procr=\proc[p']\pcmpa\procq$; and
    \item \label{prop:steplifting:item:cmerge}
      $\procp\cmergea\procq\step{\act}\procr$
        \IFF{}
      there exist actions $\acta,\actb\in\Act$ and
      processes $\proc[p'],\proc[q']\in\Proc$ such that
        $\act=\cfun(\acta,\actb)$,
        $\procp\step{\acta}\proc[p']$,
        $\procq\step{\actb}\proc[q']$,
      and
        $\procr=\proc[p']\pcmpa\proc[q']$; and
    \item \label{prop:steplifting:item:pcmp}
      $\procp\pcmpa\procq\step{\act}\procr$
        \IFF{}
      $\procp\lmergea\procq\step{\act}\procr$
      or $\procq\lmergea\procp\step{\act}\procr$
      or $\procp\cmergea\procq\step{\act}\procr$.
    \end{enumerate}
  \end{proposition}

  Let $\procp,\proc[p']\in\Proc$; we write $\procp\stepnla\proc[p']$
  if $\procp\step{\act}\proc[p']$ for some $\act\in\Acts$ and call
  $\proc[p']$ a \emph{residual} of $\procp$.
  We write $\procp\nstepnla$ if $\procp$ has no residuals.
  We denote by $\stepsnl$ the reflexive transitive closure of
  $\stepnl$.

  It is easy to see from Table~\ref{tab:tssccs} that if
  $\cptermP\step{\act}\cpterm[P']$, then $\cpterm[P']$ has fewer symbols
  than $\cptermP$. Consequently, the length of a transition sequence
  starting with a process $\eqclass{\cptermP}$ is bounded from above
  by the number of symbols in $\cptermP$.

  \begin{definition}
    The \emph{depth} $\depth{\procp}$ of an element
    $\procp\in\Proc$ is defined as
    \begin{equation*}
      \depth{\procp}=
        \max\{n\geq 0:\text{
          $\exists\proc[p_n],\dots,\proc[p_0]\in\Proc$
            s.t.\ $\procp=\proc[p_n]
                     \stepnla\cdots\stepnla
                   \proc[p_0]$}
        \}.
    \end{equation*}
    The \emph{branching degree} $\bdeg{\procp}$ of an element
    $\procp\in\Proc$ is defined as
    \begin{equation*}
      \bdeg{\procp}=
        |\{(\act,\proc[p']): \procp\step{\act}\proc[p']\}|
    \enskip.
    \end{equation*}
  \end{definition}

  For the remainder of this section, we focus on properties of
  parallel composition on \Proc{}.
  The depth of a parallel composition is the sum of the depths of its
  components.
  \begin{lemma} \label{lem:depthmergea}
  For all $\procp,\procq\in\Proc$,
    $\depth{\procp\pcmpa\procq}=\depth{\procp}+\depth{\procq}$.
  \end{lemma}
  \begin{proof}
    If
        $\procp=\proc[p_m]\stepnla\cdots\stepnla\proc[p_0]$
    and
        $\procq=\proc[q_n]\stepnla\cdots\stepnla\proc[q_0]$,
    then
    \begin{gather*}
      \procp\pcmpa\procq
        =\proc[p_m]\pcmpa\procq
           \stepnla\cdots\stepnla
         \proc[p_0]\pcmpa\procq
        =\proc[p_0]\pcmpa\proc[q_n]
           \stepnla\cdots\stepnla
         \proc[p_0]\pcmpa\proc[q_0]
    \enskip,
    \end{gather*}
    so clearly
      $\depth{\procp\pcmpa\procq}\geq\depth{\procp}+\depth{\procq}$.

    It remains to prove that
      $\depth{\procp}+\depth{\procq}\geq\depth{\procp\pcmpa\procq}$.
     We proceed by induction on the depth of $\procp\pcmpa\procq$.
    If $\depth{\procp\pcmpa\procq}=0$, then
    $(\procp\pcmpa\procq)\nstepnla$, so $\procp\nstepnla$ and
    $\procq\nstepnla$; hence $\depth{\procp}=0$ and $\depth{\procq}=0$,
    and it follows that
    $\depth{\procp\pcmpa\procq}=\depth{\procp}+\depth{\procq}$.
    Suppose that $\depth{\procp\pcmpa\procq}=n+1$.
    Then there exist $\proc[r_{n+1}],\dots,\proc[r_0]\in\Proc$ such
    that
    \begin{equation*}
      \procp\pcmpa\procq=
        \proc[r_{n+1}]\stepnla\proc[r_n]
          \stepnla\cdots\stepnla
        \proc[r_0]
    \enskip.
    \end{equation*}
    Note that $\depth{\proc[r_i]}=i$ for all $0\leq i \leq n+1$.
    Further note that the transition
      $\proc[r_{n+1}]\stepnla\proc[r_n]$ 
    cannot be the result
      $\proc[r_{n+1}]\step{\cfun(\acta,\actb)}\proc[r_n]$
    of communication between a transition
      $\procp\step{\acta}\proc[p']$
    and a transition
     $\procq\step{\actb}\proc[q']$;
    for then there would exist a longer transition sequence from
      $\procp\pcmpa\procq$,
    obtained by replacing the single transition     
    $\proc[r_{n+1}]\stepnla\proc[r_n]$ by two transitions
      $\proc[r_{n+1}]
         =\procp\pcmpa\procq
         \stepnla\proc[p']\pcmpa\procq
         \stepnla\proc[p']\pcmpa\proc[q']
         =\proc[r_n]$,
    contradicting our assumption that
    $\depth{\procp\pcmpa\procq}=n+1$.
    Hence, either $\proc[r_n]=\proc[p']\pcmpa\procq$ with
    $\proc[p]\stepnla\proc[p']$, or $\proc[r_n]=\procp\pcmpa\proc[q']$
    with $\proc[q]\stepnla\proc[q']$.
    In the first case it follows by the induction hypothesis that
      $\depth{\proc[p']}+\depth{\proc[q]}
         \geq\depth{\proc[p']\pcmpa\procq}=n$,
    so
      $\depth{\procp}+\depth{\procq}
         \geq
       \depth{\proc[p']}+\depth{\procq}+1
         \geq
       n+1=\depth{\procp\pcmpa\procq}$.
    In the second case the proof is similar.
  \qed
  \end{proof}

  According to the following lemma and
  Proposition~\ref{prop:gensound}, $\Proc$ is a commutative monoid
  with respect to $\pcmpa$, with $\nila$ as the identity element.

\newcommand{\pcmpass}{\pcmpax{2}}
\newcommand{\pcmpcom}{\pcmpax{3}}
\newcommand{\pcmpidp}{\pcmpax{4}}
  \begin{lemma} \label{lem:pcmpmonoid}
    The following equations are derivable from the axioms in
    Table~\ref{tab:axccs}:
    \begin{equation*}
      \begin{array}[t]{@{}l@{\quad}l@{\ \fequate\ }l@{}}
        \pcmpass   & (\varx\pcmp\vary)\pcmp\varz
                   & \varx\pcmp(\vary\pcmp\varz) \\
        \pcmpcom   & \varx\pcmp\vary
                   & \vary\pcmp\varx \\
        \pcmpidp   & \varx\pcmp\nil
                   & \varx
    \enskip.
      \end{array}
    \end{equation*}
  \end{lemma}

  An element $\procp\in\Proc$ is \emph{parallel prime} if
  $\procp\neq\nila$, and $\proc[p]=\procq\pcmpa\procr$ implies
  $\procq=\nila$ or $\procr=\nila$.
  Suppose that $\procp$ is an arbitrary  element of $\Proc$; a
  \emph{parallel decomposition} of $\procp$ is a finite multiset
  $[\proc[p_1],\dots,\proc[p_n]]$ of parallel primes such that
    $\procp=\proc[p_1]\pcmpa\cdots\pcmpa\proc[p_n]$.
  (The process $\nila$ has as decomposition the empty multiset, and a
  parallel prime process $\procp$ has as decomposition the singleton
  multiset $[\procp]$.)
  The following theorem is a straightforward consequence of the main
  result in \cite{LO05}.

  \begin{theorem} \label{theo:udecompmergea}
    Every element of $\Proc$ has a unique parallel decomposition.
  \end{theorem}
  \begin{proof}
    In a similar way as in \cite[Sect.\ 4]{LO05} it can be
    established that the inverse of $\stepsnla$ is a decomposition
    order on the commutative monoid $\Proc$ with respect to parallel
    composition; it then follows from \cite[Theorem~32]{LO05} that
    this commutative monoid has unique decomposition.  \qed
  \end{proof}

  The following corollary follows easily from the above unique
  decomposition result.
  \begin{corollary}[Cancellation] \label{cor:cancmergea}
    Let $\procp,\procq,\procr\in\Proc$.
    If $\procp\pcmpa\procq=\procp\pcmpa\procr$,
    then $\procq=\procr$.
  \end{corollary}

  The branching degree of a parallel composition is at least the
  branching degree of its components.
  \begin{lemma} \label{lem:bdegmergea}
    For all $\procp,\procq\in\Proc$,
      $\bdeg{\procp\pcmpa\procq}\geq\bdeg{\procp},\bdeg{\procq}$.
  \end{lemma}
  \begin{proof}
    First we prove that
      $\bdeg{\procp\pcmpa\procq}\geq\bdeg{\procq}$.
    By Proposition~\ref{prop:steplifting},
      if $\procq\step{\act}\proc[q']$,
      then $\procp\pcmpa\procq\step{\act}\procp\pcmpa\proc[q']$.
    Suppose that $\proc[q_1]$ and $\proc[q_2]$ are distinct
    processes such that $\proc[q]\step{\act}\proc[q_1]$ and
    $\proc[q]\step{\act}\proc[q_2]$.
    Then $\procp\pcmpa\procq\step{\act}\procp\pcmpa\proc[q_1]$ and 
    $\procp\pcmpa\procq\step{\act}\procp\pcmpa\proc[q_2]$.
    Since $\procp\pcmpa\proc[q_1]=\procp\pcmpa\proc[q_2]$ would imply
    $\proc[q_1]=\proc[q_2]$ by Corollary~\ref{cor:cancmergea}, it
    follows that $\procp\pcmpa\proc[q_1]$ and $\procp\pcmpa\proc[q_2]$
    are distinct. Hence $\bdeg{\procp\pcmpa\procq} \geq \bdeg{\procq}$.

    By commutativity of $\pcmpa$, it also follows
    that $\bdeg{\procp\pcmpa\procq}\geq\bdeg{\procp}$.
\qed
  \end{proof}

  We define a sequence of parallel prime processes with special
  properties that make them very suitable as tools in our proofs in
  the remainder of the paper:
  \begin{equation} \label{eq:brancherdef}
    \brancher[i] = \prefa{\silent}\nil
                     \acmp\cdots\acmp
                   \pref{\silent^i}\nil
  \qquad(i\geq 1)
  \end{equation}
  (with $\pref{\silent^i}\nil$ recursively defined by
  $\pref{\silent^i}\nil=\nil$ if $i=0$, and
  $\pref{\silent}\pref{\silent^{i-1}}\nil$ if $i>0$).
  \begin{lemma} \label{lem:brancher}
  \begin{enumerate}
  \renewcommand{\theenumi}{\roman{enumi}}
  \renewcommand{\labelenumi}{(\theenumi)}
  \item \label{lem:brancher:item:prime}
    For all $i\geq 1$, the processes $\brancher[i]$ are parallel prime.
  \item \label{lem:brancher:item:distinct}
    The processes $\brancher[i]$ are all distinct, i.e.,
    $\brancher[k]=\brancher[l]$ implies that $k=l$.
  \item \label{lem:brancher:item:bdeg}
    For all $i\geq 1$, the process $\brancher[i]$ has branching degree $i$.
  \end{enumerate}
  \end{lemma}
  \begin{proof}
  \begin{enumerate}
  \renewcommand{\theenumi}{\roman{enumi}}
  \renewcommand{\labelenumi}{(\theenumi)}
  \item
    Clearly $\brancher[i]\neq\nila$.
    Suppose $\brancher[i]=\procp\pcmpa\procq$; to prove that
    $\brancher[i]$ is parallel prime, we need to establish that either
    $\procp=\nil$ or $\procq=\nil$.
    Note that $\procp\pcmp\procq\step{\silent}\nil$.
    There do not exist actions $\acta$ and $\actb$ and processes
    $\proc[p']$ and $\proc[q']$ such that $\cfun(\acta,\actb)=\silent$
    and $\proc[p']\pcmpa\proc[q']$, for then also
    $\procp\pcmpa\procq\step{\acta}\proc[p']\pcmpa\proc[q]$, quod
    non. Therefore, according to Proposition~\ref{prop:steplifting},
    there are only two cases to consider:
    \begin{enumerate}
    \item If there exists $\proc[p']$ such that
      $\procp\step{\silent}\proc[p']$ and
      $\proc[p']\pcmpa\proc[q]=\nil$,
      then it follows by Lemma~\ref{lem:depthmergea} that
      $\depth{\proc[q]}=0$, and hence $\procq=\nil$.
    \item If there exists $\proc[q']$ such that
      $\procq\step{\silent}\proc[q']$ and
      $\procp\pcmpa\proc[q']=\nil$,
      then it follows by Lemma~\ref{lem:depthmergea} that
      $\depth{\procp}=0$, and hence $\procp=\nil$.
    \end{enumerate}
  \item
    If $\brancher[k]=\brancher[l]$, then
    $k=\depth{\brancher[k]}=\depth{\brancher[l]}=l$.
  \item
    On the one hand,
      $\brancher[i]\step{\silent}\pref{\silent^{j}}\nil$
    for all $0\leq j < i$ and
    $\pref{\silent^{k}}\nil=\pref{\silent^l}\nil$ implies $k=l$ for
    all $0\leq k,l < i$, so $\bdeg{\brancher[i]}$ is at least $i$.
    On the other hand, if $\brancher[i]\step{\act}\proc$, then
    $\act=\silent$ and $\proc=\pref{\silent^{j}}\nil$ for some $0\leq
    j < i$, so $\bdeg{\brancher[i]}$ is at most $i$.
\qed
  \end{enumerate}
  \end{proof}

  \section{An Equational Base for $\ProcFM$} \label{sec:interleaving}

  In this section, we prove that an equational base for $\ProcFM$ is
  obtained if the axiom
  \begin{equation*}
    \fmax{}\quad\varx\cmerge\vary\ \fequate\ \nil
  \end{equation*}
  is added to the set of axioms generated by the axiom schemata in
  Table~\ref{tab:axccs}. The resulting equational base is
  finite if $\Act$ is finite.
  Henceforth, whenever we write $\ptermP\feqfm\ptermQ$, we mean that
  the equation $\ptermP\fequate\ptermQ$ is derivable from
  the axioms in Table~\ref{tab:axccs} and the axiom \fmax{}.

  \begin{proposition} \label{prop:fmsound}
    For all process terms $\ptermP$ and $\ptermQ$, if
    $\ptermP\feqfm\ptermQ$, then $\ptermP\bisimilarfm\ptermQ$.
  \end{proposition}

  To prove that adding $\fmax{}$ to the axioms in
  Table~\ref{tab:axccs} suffices to obtain an equational base for
  $\ProcFM$, we need to establish that $\ptermP\bisimilarfm\ptermQ$
  implies $\ptermP\feqfm\ptermQ$ for all process terms $\ptermP$ and
  $\ptermQ$.
  First, we identify a set of normal forms $\fNFTERMS$ such that every
  process term $\ptermP$ can be rewritten to a normal form by means of
  the axioms.
  
  \begin{definition} \label{def:fnf}
    The set $\fNFTERMS$ of \emph{\fmax{}-normal forms} is generated
    by the following grammar:
    \begin{equation*}
       \ptermN\ ::=\
         \nil\ \mid\
         \ptermN\acmp\ptermN\ \mid\
         \pref{\act}\ptermN\ \mid\
         \varx\lmerge\ptermN
   \enskip,
   \end{equation*}
    with $\varx\in\VAR$, and $\act\in\Acts$.
  \end{definition}

  \begin{lemma} \label{lem:fnf}
    For every process term $\ptermP$ there is an \fmax{}-normal form
    $\ptermN$ such that $\ptermP\feqfm\ptermN$ and
    $\height{\ptermP}\geq\height{\ptermN}$.
  \end{lemma}
  \begin{proof}
  Recall that $\height{\ptermP}$ denotes the height of $\ptermP$ (see
  Definition~\ref{def:height}). In this proof we also use another
  syntactic measure on $\ptermP$: the \emph{length} of $\ptermP$,
  denoted $\length{\ptermP}$, is the number of symbols occurring in
  $\ptermP$.
  Define a partial order $\prec$ on process terms by
  $\ptermP\prec\ptermQ$ if the pair
    $(\height{\ptermP},\length{\ptermP})$
  is less than the pair
    $(\height{\ptermQ},\length{\ptermQ})$
  in the lexicographical order on $\N\times\N$;
  i.e., $\ptermP\prec\ptermQ$ if $\height{\ptermP}<\height{\ptermQ}$
  or $\height{\ptermP}=\height{\ptermQ}$ and
  $\length{\ptermP}<\length{\ptermQ}$.
  It is well-known that the lexicographical order on
  $\N\times\N$, and hence the order $\prec$ on process
  terms, is well-founded; so we may use $\prec$-induction.

  The remainder of the proof consists of a case distinction on the
  syntactic forms that $\ptermP$ may take.
  \begin{enumerate}
  \item
    If $\ptermP$ is a variable, say $\ptermP=\varx$, then
      $\ptermP\fequate\varx\lmerge\nil$ by \lmnil{};
    the process term $\varx\lmerge\nil$ is an \fmax{}-normal form and
      $\height{\ptermP}
         =\height{\varx}
         =\height{\varx}+0
         =\height{\varx\lmerge\nil}$.

  \item
    If $\ptermP=\nil$, then $\ptermP$ is an \fmax{}-normal form.

  \item
    If $\ptermP=\pref{\act}\pterm[P']$, then, since
    $\height{\pterm[P']}<\height{\ptermP}$, it holds that
    $\pterm[P']\prec\ptermP$, and hence by the induction hypothesis
    there exists an \fmax{}-normal form $\pterm[N]$ such that
      $\pterm[P']\feqfm\pterm[N]$
    and
      $\height{\pterm[P']}\geq\height{\pterm[N]}$.
    Then $\pref{\act}\pterm[N]$ is an \fmax{}-normal form such that
      $\pterm[P]\feqfm\pref{\act}\pterm[N]$
    and
      $\height{\pterm[P]}\geq\height{\pref{\act}\pterm[N]}$.

  \item
    If $\ptermP=\pterm[P_1]\acmp\pterm[P_2]$, then, since
    $\height{\pterm[P_1]},\height{\pterm[P_2]}\leq\height{\ptermP}$
    and $\length{\pterm[P_1]},\length{\pterm[P_2]}<\length{\ptermP}$,
    it holds that $\pterm[P_1],\pterm[P_2]\prec\ptermP$, and hence by
    the induction hypothesis there exist \fmax{}-normal forms
    $\pterm[N_1]$ and $\pterm[N_2]$ such that
      $\pterm[P_1]\feqfm\pterm[N_1]$,
      $\pterm[P_2]\feqfm\pterm[N_2]$,
      $\height{\pterm[P_1]}\geq\height{\pterm[N_1]}$
    and
      $\height{\pterm[P_2]}\geq\height{\pterm[N_2]}$.
    Then $\pterm[N_1]\acmp\pterm[N_2]$ is an \fmax{}-normal form such
    that
      $\ptermP\feqfm\pterm[N_1]\acmp\pterm[N_2]$
    and
      $\height{\ptermP}\geq\height{\pterm[N_1]\acmp\pterm[N_2]}$.

  \item \label{case:flmerge}
    If $\ptermP=\pterm[Q]\lmerge\pterm[R]$, then, since
    $\height{\pterm[Q]}\leq\height{\ptermP}$ and
    $\length{\pterm[Q]}<\length{\ptermP}$, it holds that
    $\pterm[Q]\prec\ptermP$, and hence by the induction hypothesis
    and  Lemma~\ref{lem:simplify} there exists a collection
    $\pterm[S_1],\dots,\pterm[S_n]$ of simple \fmax{}-normal forms
    such that
    $\pterm[Q]\feqfm\sum_{i=1}^n\pterm[S_i]$ and
      $\height{\pterm[Q]}\geq\height{\pterm[S_i]}$
        for all $i=1,\dots,n$.
    If $n=0$, then
      $\pterm\feqfm\nil\lmerge\pterm[R]\fequate\nil$
    by \nillm{}, and clearly $\height{\ptermP}\geq\height{\nil}$.
    Otherwise, by \aclm{}
    \begin{equation*}
      \ptermP \feqfm \sum_{i=1}^n (\pterm[S_i]\lmerge\pterm[R])
    \enskip.
    \end{equation*}
    So it remains to show, for all $i=1,\dots,n$, that
    $\pterm[S_i]\lmerge\pterm[R]$ is provably equal to an appropriate
    \fmax{}-normal form.
    We distinguish cases according to the syntactic form of
    $\pterm[S_i]$:
    \begin{enumerate}
    \item If $\pterm[S_i]=\pref{\act}\pterm[N_i']$, with
      $\pterm[N_i']$ an \fmax{}-normal form, then by \preflm{}
      \begin{equation*}
        \pterm[S_i]\lmerge\ptermR
          \fequate\pref{\act}(\pterm[N_i']\pcmp\ptermR)
      \enskip.
      \end{equation*}
      Since
        $\height{\pterm[N_i']}<\height{\pterm[S_i]}\leq\height{\ptermQ}$,
      it holds that $\pterm[N_i']\pcmp\ptermR \prec\ptermP$
      and hence by the induction hypothesis there exists an
      \fmax{}-normal form $\pterm[N_i]$ such that
        $\pterm[N_i']\pcmp\ptermR\feqfm\pterm[N_i]$
      and
        $\height{\pterm[N_i']\pcmp\ptermR}\geq\height{\pterm[N_i]}$.
      Clearly,
        $\pref{\act}\pterm[N_i]$
      is an \fmax{}-normal form such that
        $\pterm[S_i]\lmerge\pterm[R]
           \feqfm\pref{\act}\pterm[N_i]$
      and
        $\height{\pterm[S_i]\lmerge\pterm[R]}\geq
           \height{\pref{\act}\pterm[N_i]}$.
    \item If $\pterm[S_i]=\varx\lmerge\pterm[N_i']$, with
      $\pterm[N_i']$ an \fmax{}-normal form, then by \lmlm{}
      \begin{equation*}
        (\varx\lmerge\pterm[N_i'])\lmerge\ptermR
          \fequate\varx\lmerge(\pterm[N_i']\pcmp\ptermR)
      \enskip.
      \end{equation*}
      Note that $\height{\varx}=1$, so
        $\height{\pterm[N_i']}<\height{\pterm[S_i]}\leq\height{\ptermQ}$.
      It follows that $\pterm[N_i']\pcmp\ptermR\prec\ptermP$, and
      hence by the induction hypothesis there exists an \fmax{}-normal
      form $\pterm[N_i]$ such that
        $\pterm[N_i']\pcmp\ptermR\feqfm\pterm[N_i]$
      and
        $\height{\pterm[N_i']\pcmp\ptermR}\geq\height{\pterm[N_i]}$.
      Clearly,
        $\varx\lmerge\pterm[N_i]$
      is an \fmax{}-normal form such that
        $\pterm[S_i]\lmerge\pterm[R]
           \feqfm\varx\lmerge\pterm[N_i]$
      and
        $\height{\pterm[S_i]\lmerge\pterm[R]}\geq
           \height{\varx\lmerge\pterm[N_i]}$.
    \end{enumerate}

  \item \label{case:fcmerge}
    If $\ptermP=\pterm[Q]\cmerge\pterm[R]$, then $\ptermP\feqfm\nil$
    according to the axiom \fmax{} and clearly $\height{\ptermP}\geq
    \height{\nil}$.

  \item
    If $\ptermP=\ptermQ\pcmp\ptermR$, then
      $\ptermP
         \fequate(\ptermQ\lmerge\ptermR
                   \acmp
                 \ptermR\lmerge\ptermQ)
                   \acmp
                 \ptermQ\cmerge\ptermR
         \feqfm\ptermQ\lmerge\ptermR\acmp\ptermR\lmerge\ptermQ$
    by the axioms \pcmplmcm{}, \fmax{} and \acnil{}.
    We can now proceed as in case~\ref{case:flmerge} to show that for
    $\ptermQ\lmerge\ptermR$ and $\ptermR\lmerge\ptermQ$ there exist
    \fmax{}-normal forms $\pterm[N_1]$ and $\pterm[N_2]$,
    respectively, such that
      $\ptermQ\lmerge\ptermR\feqfm\pterm[N_1]$,
      $\ptermR\lmerge\ptermQ\feqfm\pterm[N_2]$,
      $\height{\ptermQ\lmerge\ptermR}\geq\height{\pterm[N_1]}$
    and
      $\height{\ptermR\lmerge\ptermQ}\geq\height{\pterm[N_2]}$.
    Then $\pterm[N_1]\acmp\pterm[N_2]$ is an \fmax{}-normal form such
    that
      $\ptermP\feqfm\pterm[N_1]\acmp\pterm[N_2]$
    and
      $\height{\ptermP}
         \geq\height{\pterm[N_1]\acmp\pterm[N_2]}$.
  \qed
  \end{enumerate}
  \end{proof}

  It remains to prove that for every two \fmax{}-normal forms
  $\pterm[N_1]$ and $\pterm[N_2]$ there exists a \emph{distinguishing
  valuation}, i.e., a valuation $\fvalstar$ such that if
  $\pterm[N_1]$ and $\pterm[N_2]$ are \emph{not} provably equal, then
  the $\fvalstar$-interpretations of $\pterm[N_1]$ and $\pterm[N_2]$
  are distinct. Stating it contrapositively, for every two
  \fmax{}-normal forms $\pterm[N_1]$ and $\pterm[N_2]$, it suffices to
  establish the existence of a valuation
    $\valstar:\VAR\rightarrow\Proc$
  such that
  \begin{equation} \label{eq:fmgoal}
     \text{if}\
       \fevalstar{\pterm[N_1]}=\fevalstar{\pterm[N_2]},\ 
     \text{then}\ \pterm[N_1]\feqfm\pterm[N_2].
  \end{equation}

  The idea is to use a valuation $\fvalstar$ that assigns processes to
  variables in such a way that much of the original syntactic
  structure of $\pterm[N_1]$ and $\pterm[N_2]$ can be recovered by
  analysing the behaviour of $\fevalstar{\pterm[N_1]}$ and
  $\fevalstar{\pterm[N_2]}$.  To recognize variables, we shall use the
  special processes $\brancher[i]$ ($i\geq 1$) defined in Eqn.\
  \eqref{eq:brancherdef} on p.\ \pageref{eq:brancherdef}. Recall that
  the processes $\brancher[i]$ have branching degree $i$.  We are
  going to assign to every variable a distinct process $\brancher[i]$.
  By choosing $i$ larger than the maximal `branching degrees'
  occurring in $\pterm[N_1]$ and $\pterm[N_2]$, the behaviour
  contributed by an instantiated variable is distinguished from
  behaviour already present in the \fmax{}-normal forms themselves.

  \begin{definition} \label{def:fwidth} We define the \emph{width}
    $\fwidth{\ptermN}$ of an \fmax{}-normal form $\ptermN$ as follows:
    \begin{enumerate}
    \renewcommand{\theenumi}{\roman{enumi}}
    \renewcommand{\labelenumi}{(\theenumi)}
    \item \label{def:fwidth:item:nil}
      if $\ptermN=\nil$, then $\fwidth{\ptermN}=0$;
    \item \label{def:fwidth:item:acmp}
      if $\ptermN=\pterm[N_1]\acmp\pterm[N_2]$, then
        $\fwidth{\pterm[N]}=\fwidth{\pterm[N_1]}+\fwidth{\pterm[N_2]}$; 
    \item \label{def:fwidth:item:pref}
      if $\ptermN=\pref{\act}\pterm[N']$, then
      $\fwidth{\ptermN}=\max(\fwidth{\pterm[N']},1)$;
    \item if $\ptermN=\varx\lmerge\pterm[N']$, then
      $\fwidth{\ptermN}=\max(\fwidth{\pterm[N']},1)$.
    \end{enumerate}
  \end{definition}

  The valuation $\fvalstar$ that we now proceed to define is
  parametrised with a natural number $\FMWidth$; in
  Theorem~\ref{theo:fnfdistinction} we shall prove that it serves as a
  distinguishing valuation (i.e., satisfies Eqn.\ \eqref{eq:fmgoal})
  for all \fmax{}-normal forms $\pterm[N_1]$ and $\pterm[N_2]$ such
  that $\fwidth{\pterm[N_1]},\fwidth{\pterm[N_2]}\leq\FMWidth$.
  Let $\fpvarnum{\_}$ denote an \emph{injective} function
  \begin{equation*}
    \fpvarnum{\_}:\VAR\rightarrow\{n\in\N: n>\FMWidth\}
  \end{equation*}
  that associates with every variable a unique natural number greater
  than $\FMWidth$.
  We define the valuation $\fvalstar:\VAR\rightarrow\Proc$ for all
  $\varx\in\VAR$ by
  \begin{equation*}
    \fvalstar(\varx) = \pref{\silent}
                        \brancher[\fpvarnum{\varx}]
  \enskip.
  \end{equation*}
  The $\silent$-prefix is to ensure the following property for
  \emph{all} normal forms (not just $\pterm[N_1]$ and $\pterm[N_2]$).
  \begin{lemma} \label{lem:fbdegnf}
    For every \fmax{}-normal form $\ptermN$,
    the branching degree of $\fevalstar{\ptermN}$ is at most
    $\fwidth{\ptermN}$.
  \end{lemma}
  \begin{proof}
    Structural induction on $\ptermN$.
  \qed
  \end{proof}

  \begin{lemma} \label{lem:frs}
    Let $\pterm[S]$ be a simple \fmax{}-normal form, let
    $\act\in\Acts$, and let $\procp$ be a process such that
      $\fevalstar{\pterm[S]}\step{\act}\procp$.
    Then the following statements hold:
    \begin{enumerate}
    \renewcommand{\theenumi}{\roman{enumi}}
    \renewcommand{\labelenumi}{(\theenumi)}
    \item \label{lem:frs:item:pref}
      if $\ptermS=\pref{\act[\beta]}\pterm[N]$,
      then $\act=\act[\beta]$ and $\proc[p]=\fevalstar{\pterm[N]}$;
    \item \label{lem:frs:item:varlmerge}
      if $\ptermS=\varx\lmerge\pterm[N]$,
      then
        $\act=\silent$
      and
        $\proc[p]=\brancher[\fpvarnum{\varx}]
                    \pcmp\fevalstar{\ptermN}$.
    \end{enumerate}
  \end{lemma}

  An important property of $\fvalstar$ is that it allows us to
  distinguish the different types of simple \fmax{}-normal forms by
  classifying their residuals according to the number of parallel
  components with a branching degree that exceeds $\FMWidth$.  Let us
  say that a process $\procp$ is of \emph{type $n$} ($n\geq 0$) if its
  unique parallel decomposition contains precisely $n$ parallel prime
  components with a branching degree $> \FMWidth$.
  \begin{corollary} \label{cor:frt}
    Let $\ptermS$ be a simple \fmax{}-normal form such that
    $\fwidth{\pterm[S]}\leq \FMWidth$.
    \begin{enumerate}
    \renewcommand{\theenumi}{\roman{enumi}}
    \renewcommand{\labelenumi}{(\theenumi)}
    \item \label{cor:frt:item:pref}
      If $\ptermS=\pref{\act}\ptermN$, then the unique residual
      $\fevalstar{\pterm[N]}$ of $\fevalstar{\ptermS}$ is of type $0$.
    \item \label{cor:frt:item:varlmerge}
      If $\ptermS=\varx\lmerge\ptermN$, then the unique residual
      $\brancher[\fpvarnum{\varx}]\pcmp\fevalstar{\ptermN}$ of
      $\fevalstar{\ptermS}$ is of type $1$.
    \end{enumerate}
  \end{corollary}
  \begin{proof}
    On the one hand, by Lemma~\ref{lem:fbdegnf}, in both cases
    $\fevalstar{\pterm[N]}$ has a branching degree of at most
    $\fwidth{\ptermN}\leq\fwidth{\ptermS}\leq\FMWidth$, and 
    hence, by Lemma~\ref{lem:bdegmergea}, its unique parallel
    decomposition cannot contain parallel prime components with a
    branching degree that exceeds $\FMWidth$. On the other hand, by
    Lemmas~\ref{lem:brancher}(\ref{lem:brancher:item:prime}) and
    \ref{lem:brancher}(\ref{lem:brancher:item:bdeg}), the process
      $\brancher[\fpvarnum{\varx}]$
    is parallel prime and has a branching degree that exceeds
    $\FMWidth$.
    So $\fevalstar{\ptermN}$ is of type $0$, and
    $\brancher[\fpvarnum{\varx}]\pcmp\fevalstar{\ptermN}$ is of type
    $1$.
  \qed
  \end{proof}

  \begin{theorem} \label{theo:fnfdistinction}
    For every two \fmax{}-normal forms $\pterm[N_1]$, $\pterm[N_2]$
    such that $\fwidth{\pterm[N_1]},\fwidth{\pterm[N_2]}\leq W$
    it holds that
      $\hevalstar{\pterm[N_1]}=\hevalstar{\pterm[N_2]}$
    only if
      $\pterm[N_1]\fequate\pterm[N_2]$ modulo \accom{}--\acnil{}.
  \end{theorem}
  \begin{proof}
    By Lemma~\ref{lem:simplify} we may assume that $\pterm[N_1]$ and
    $\pterm[N_2]$ are summations of collections of simple \fmax{}-normal
    forms.
    We assume $\fevalstar{\pterm[N_1]}=\fevalstar{\pterm[N_2]}$ and
    prove that then $\pterm[N_1]\fequate\pterm[N_2]$ modulo
    \accom{}--\acnil{}, by induction on the sum of the heights of
    $\pterm[N_1]$ and $\pterm[N_2]$.

    We first prove that for every syntactic summand $\pterm[S_1]$ of
    $\pterm[N_1]$ there is a syntactic summand $\pterm[S_2]$ of
    $\pterm[N_2]$ such that $\pterm[S_1]\fequate\pterm[S_2]$ modulo 
    \accom{}--\acnil{}.
    To this end, let $\pterm[S_1]$ be an arbitrary syntactic summand
    of $\pterm[N_1]$; we distinguish cases according to the syntactic
    form of $\pterm[S_1]$.
    \begin{enumerate}
    \item Suppose $\pterm[S_1]=\pref{\act}\pterm[N_1']$;
       then
         $\fevalstar{\pterm[S_1]}\step{\act}\fevalstar{\pterm[N_1']}$.
       Hence, since $\hevalstar{\pterm[N_1]}=\hevalstar{\pterm[N_2]}$,
       there exists a syntactic summand $\pterm[S_2]$ of $\pterm[N_2]$
       such that
         $\fevalstar{\pterm[S_2]}\step{\act}\fevalstar{\pterm[N_1']}$.
       By Lemma~\ref{lem:fbdegnf} the branching degree of
       $\fevalstar{\pterm[N_1']}$ does not exceed $\FMWidth$, so
       $\fevalstar{\pterm[S_2]}$ has a residual of type $0$, and
       therefore, by Corollary~\ref{cor:frt}, there exist
       $\act[\beta]\in\Acts$ and an \fmax{}-normal form $\pterm[N_2']$
       such that 
         $\pterm[S_2]=\pref{\act[\beta]}\pterm[N_2']$.
       Moreover, since
         $\fevalstar{\pterm[S_2]}\step{\act}\fevalstar{\pterm[N_1']}$,
       it follows by
       Lemma~\ref{lem:frs}(\ref{lem:frs:item:pref})
       that $\act=\act[\beta]$ and
       $\fevalstar{\pterm[N_1']}=\fevalstar{\pterm[N_2']}$.
       Hence, by the induction hypothesis, we conclude that
         $\pterm[N_1']\fequate\pterm[N_2']$ modulo \accom{}--\acnil{},
       so $\pterm[S_1]=\pref{\act}\pterm[N_1']
             \fequate\pref{\act[\beta]}\pterm[N_2']=\pterm[S_2]$.
    \item Suppose $\pterm[S_1]=\varx\lmerge\pterm[N_1']$;
      then
        $\fevalstar{\pterm[S_1]}
           \step{\silent}
             \brancher[\fpvarnum{\varx}]\pcmp\fevalstar{\pterm[N_1']}$.
      Hence, since $\fevalstar{\pterm[N_1]}=\hevalstar{\pterm[N_2]}$,
      there exists a summand $\pterm[S_2]$ of $\pterm[N_2]$ such that
        $\fevalstar{\pterm[S_2]}
           \step{\silent}
             \brancher[\fpvarnum{\varx}]\pcmp\fevalstar{\pterm[N_1']}$.
      Since $\pterm[S_2]$ has a residual of type 1, by
      Corollary~\ref{cor:frt} there exist a variable
      $\vary$ and an \fmax{}-normal form $\pterm[N_2']$ such that
        $\pterm[S_2]=\vary\lmerge\pterm[N_2']$.
      Now, since
        $\fevalstar{S_2}
           \step{\silent}
             \brancher[\fpvarnum{\varx}]\pcmp\fevalstar{\pterm[N_1']}$,
      it follows by
      Lemma~\ref{lem:frs}(\ref{lem:frs:item:varlmerge})
      that
      \begin{equation} \label{eq:fparalleleq1}
         \brancher[\fpvarnum{\varx}]\pcmp\fevalstar{\pterm[N_1']}
       = \brancher[\fpvarnum{\vary}]\pcmp\fevalstar{\pterm[N_2']}
      \enskip.
      \end{equation}
      Since $\fevalstar{\pterm[N_1']}$ and $\fevalstar{\pterm[N_2']}$
      are of type $0$, we have that the unique decomposition
      of $\fevalstar{\pterm[N_1']}$
        (see Theorem~\ref{theo:udecompmergea})
      does not contain
      $\brancher[\fpvarnum{\vary}]$ and the unique decomposition of
      $\fevalstar{\pterm[N_2']}$ does not contain
      $\brancher[\fpvarnum{\varx}]$.
      Hence, from \eqref{eq:fparalleleq1} it follows that
        $\brancher[\fpvarnum{\varx}]=\brancher[\fpvarnum{\vary}]$
      and
        $\fevalstar{\pterm[N_1']}=\fevalstar{\pterm[N_2']}$.
      From the former we conclude,
      by Lemma~\ref{lem:brancher}(\ref{lem:brancher:item:distinct})
      and the injectivity of $\hpvarnum{.}$, that
      $\varx=\vary$ and from the latter we conclude by the induction
      hypothesis that
        $\pterm[N_1']\fequate\pterm[N_2']$ modulo \accom{}--\acnil{}.
      So $\pterm[S_1]=\varx\lmerge\pterm[N_1']
            \fequate
          \vary\lmerge\pterm[N_2']=\pterm[S_2]$.
    \end{enumerate}
    We have established that every syntactic summand of $\pterm[N_1]$
    is provably equal to a syntactic summand of
    $\pterm[N_2]$. Similarly, it follows that every syntactic summand
    of $\pterm[N_2]$ is provably equal to a syntactic summand of 
    $\pterm[N_1]$. Hence, modulo \accom{}--\acnil{},
      $\pterm[N_1]
         \fequate\pterm[N_1]\acmp\pterm[N_2]\fequate
           \pterm[N_2]$,
    so the proof of the theorem is complete.
  \qed
  \end{proof}

  Note that it follows from the preceding theorem that there exists a
  distinguishing valuation for \emph{every} pair of \fmax{}-normal
  forms $\pterm[N_1]$ and $\pterm[N_2]$ that are distinct modulo
  \accom{}--\acnil{}; it is obtained by instantiating the parameter
  $\FMWidth$ in the definition of $\fvalstar$ with a sufficiently
  large value. Hence, we get the following corollary.

  \begin{corollary} \label{cor:fmcomplete}
    For all process terms $\ptermP$ and $\ptermQ$,
      $\ptermP\feqfm\ptermQ$
    if, and only if,
      $\ptermP\bisimilarfm\ptermQ$,
    and hence the axioms generated by the schemata in
    Table~\ref{tab:axccs} together with the axiom \fmax{} consitute an
    equational base for $\ProcFM{}$. 
  \end{corollary}
  \begin{proof}
    The implication from left to right is
    Proposition~\ref{prop:fmsound}.
    To prove the implication from right to left, suppose
    $\ptermP\bisimilarfm\ptermQ$. Then, by Lemma~\ref{lem:fnf} there
    exist \fmax{}-normal forms $\pterm[N_1]$ and $\pterm[N_2]$ such
    that $\ptermP\feqfm\pterm[N_1]$ and $\ptermQ\feqfm\pterm[N_2]$;
    from $\ptermP\bisimilarfm\ptermQ$ we conclude by
    Proposition~\ref{prop:fmsound} that
      $\pterm[N_1]\bisimilarfm\pterm[N_2]$.
    Now choose $\FMWidth$ large enough such that
      $\fwidth{\pterm[N_1]},\fwidth{\pterm[N_2]}\leq W$.
    From $\pterm[N_1]\bisimilarfm\pterm[N_2]$ it follows that
    $\hevalstar{\pterm[N_1]}=\hevalstar{\pterm[N_2]}$, and hence, by
    Theorem~\ref{theo:fnfdistinction} $\pterm[N_1]\fequate\pterm[N_2]$.
    We may therefore conclude that
      $\pterm[P]
         \feqfm
       \pterm[N_1]
         \fequate
       \pterm[N_2]
         \feqfm
       \pterm[Q]$.
  \qed
  \end{proof}

  \begin{corollary}
    The equational theory of $\ProcFM{}$ is decidable.
  \end{corollary}
  \begin{proof}
    From the proof of Lemma~\ref{lem:fnf} it is easy to see that
    there exists an effective procedure that associates with every
    process term a provably equivalent \fmax{}-normal.
    Furthermore, from Definition~\ref{def:fwidth} it is clear that
    every \fmax{}-normal form has an effectively computable width.
    We now sketch an effective procedure that decides whether a
    process equation
      $\ptermP\fequate\ptermQ$
    is valid:
    \begin{enumerate}
    \item Compute \fmax{}-normal forms $\pterm[N_1]$ and $\pterm[N_2]$
      such that $\ptermP\feqfm\pterm[N_1]$ and $\ptermQ\feqfm\pterm[N_2]$.
    \item Compute $\fwidth{\pterm[N_1]}$ and $\fwidth{\pterm[N_2]}$
      and define $\FMWidth$ as their maximum.
    \item Determine the (finite) set $\VAR[V']$ of variables
      occurring in $\pterm[N_1]$ and $\pterm[N_2]$; define an
      injection
        $\fpvarnum{.}:\VAR[V']\rightarrow\{n\in\N:n>\FMWidth\}$,
      and a substitution
        $\fvalstar:\VAR[V']\rightarrow\CPTERMS$
      that assigns to a variable $\varx$ in $\VAR[V']$ the closed
      process term $\pref{\silent}\brancher[\fpvarnum{\varx}]$.
      (We may interpret Eqn.~\eqref{eq:brancherdef} as
      defining a sequence of closed process terms instead of a
      sequence of processes.) 
    \item Let $\pterm[N_1^{\fvalstar}]$ and $\pterm[N_2^{\fvalstar}]$
      be the results from applying $\fvalstar$ to $\pterm[N_1]$ and
      $\pterm[N_2]$, respectively.
    \item Determine if the closed process terms
      $\pterm[N_1^{\fvalstar}]$ and $\pterm[N_2^{\fvalstar}]$ are
      bisimilar;
      if they are, then the process equation
        $\ptermP\fequate\ptermQ$
      is valid in $\ProcFM$, and otherwise it is not.
\qed
    \end{enumerate}
  \end{proof}

  \section{An Equational Base for \ProcCCS{}}
  \label{sec:ccsparallel}

  We now consider the algebra \ProcCCS{}. Note that if $\Act$ happens
  to be the empty set, then \ProcCCS{} satisfies the axiom \fmax{},
  and it is clear from the proof in the previous section that the
  axioms generated by the axiom schemata in Table~\ref{tab:axccs}
  together with \fmax{} in fact constitute a finite equational base
  for \ProcCCS{}. We therefore proceed with the assumption that $\Act$
  is nonempty, and prove that an equational base for $\ProcCCS$
  is then obtained if we add the axiom
  \begin{equation*}
    \hsax{}\quad\varx\cmerge(\vary\cmerge\varz) \fequate\ \nil
  \end{equation*}
  to the set of axioms generated by the axiom schemata in
  Table~\ref{tab:axccs}. Again, the resulting equational base is
  finite if the set $\Act$ is finite.
  Henceforth, whenever we write $\ptermP\feqhs\ptermQ$, we
  mean that the equation $\ptermP\fequate\ptermQ$ is derivable from
  the axioms in Table~\ref{tab:axccs} and the axiom \hsax{}.

  \begin{proposition} \label{prop:ccssound}
    For all process terms $\ptermP$ and $\ptermQ$, if
    $\ptermP\feqhs\ptermQ$, then $\ptermP\bisimilarccs\ptermQ$.
  \end{proposition}

  We proceed to adapt the proof presented in the previous section to
  establish the converse of Proposition~\ref{prop:ccssound}.
  Naturally, with \hsax{} instead of \fmax{} not every occurrence of
  $\cmerge$ can be eliminated from process terms, so the first thing
  we need to do is to adapt the notion of normal form.

  \begin{definition}
    The set $\hNFTERMS$ of \emph{\hsax{}-normal forms} is generated
    by the following grammar:
    \begin{equation*}
       \pterm[N]\ ::=\
         \nil\ \mid\
         \ptermN\acmp\ptermN\ \mid\
         \pref{\act}\ptermN\ \mid\
         \varx\lmerge\ptermN\ \mid\
         (\varx\cmerge\acta)\lmerge\ptermN\ \mid
         (\varx\cmerge\vary)\lmerge\ptermN
    \enskip,
    \end{equation*}
    with $\varx,\vary\in\VAR$, $\act\in\Acts$ and $\acta\in\Act$.
  \end{definition}

   In the proof that every process term is provably equal to an
   \hsax{}-normal form, we use the following derivable equation.
\newcommand{\cmsilent}{\cmergeax{9}}
  \begin{lemma} \label{lem:silentcom}
    The following equation is derivable from the axioms in
    Table~\ref{tab:axccs} and the axiom \hsax{}:
    \begin{equation*}
    \begin{array}[t]{@{}l@{\quad}l@{\ \feqhs\ }l@{}}
        \cmsilent & \pref{\silent}\varx\cmerge\vary
                  & \nil
    \enskip.
    \end{array}
    \end{equation*}
  \end{lemma}
  \begin{proof}
    Let $\acta\in\Act$; then
    \begin{alignat*}{2}
      \pref{\silent}\varx\cmerge\vary
    & \feqhs
      \pref{\silent}(\varx\pcmp\nil)\cmerge\vary
    &&\quad\text{by \pcmpidp{} (see Lemma~\ref{lem:pcmpmonoid})} \\
    & \feqhs
      (\pref{\acta}\varx\cmerge\pref{\coact}\nil)\cmerge\vary
    &&\quad\text{by \prefcmd{}}\\
    & \feqhs
      \nil
    &&\quad\text{by \hsax{}.}
    \end{alignat*}
\qed
  \end{proof}

  \begin{lemma} \label{lem:hnf}
    For every process term $\ptermP$ there exists an \hsax{}-normal
    form $\ptermN$ such that $\ptermP\feqhs\ptermN$ and
    $\height{\ptermP}\geq\height{\ptermN}$.
  \end{lemma}
  \begin{proof}
    As in the proof of Lemma~\ref{lem:fnf} we proceed by
    $\prec$-induction and do a case distinction on the
    syntactic form of $\ptermP$. For the first four cases ($\ptermP$
    is a variable, $\ptermP=\nil$, $\ptermP=\pref{\act}\pterm[P']$ and
    $\ptermP=\pterm[P_1]\acmp\pterm[P_2]$) the proofs are identical to
    those in Lemma~\ref{lem:fnf}, so they are omitted.

  \begin{enumerate}
\addtocounter{enumi}{4}
  \item \label{case:hlmerge}
    If $\ptermP=\pterm[Q]\lmerge\pterm[R]$, then, since
    $\height{\pterm[Q]}\leq\height{\ptermP}$ and
    $\length{\pterm[Q]}<\length{\ptermP}$, it holds that
    $\pterm[Q]\prec\ptermP$, and hence by the induction hypothesis
    and  Lemma~\ref{lem:simplify} there exists a collection
    $\pterm[S_1],\dots,\pterm[S_n]$ of simple \hsax{}-normal forms
    such that $\pterm[Q]\feqhs\sum_{i=1}^n\pterm[S_i]$ and
      $\height{\pterm[Q]}\geq\height{\pterm[S_i]}$
        for all $i=1,\dots,n$.
    If $n=0$, then
      $\ptermP\feqhs\nil\lmerge\pterm[R]\fequate\nil$
    by \nillm{}, and clearly $\height{\ptermP}\geq\height{\nil}$.
    Otherwise, by \aclm{}
    \begin{equation*}
      \ptermP \feqhs \sum_{i=1}^n (\pterm[S_i]\lmerge\pterm[R])
    \enskip,
    \end{equation*}
    so it remains to show, for all $i=1,\dots,n$, that
    $\pterm[S_i]\lmerge\pterm[R]$ is provably equal to an appropriate
    \hsax{}-normal form.
    We distinguish cases according to the syntactic form of
    $\pterm[S_i]$:
    \begin{enumerate}
    \item If $\pterm[S_i]=\pref{\act}\pterm[N_i']$ (with
      $\pterm[N_i']$ an \hsax{}-normal form), then by \preflm{}
      \begin{equation*}
        \pterm[S_i]\lmerge\ptermR
          \feqhs\pref{\act}(\pterm[N_i']\pcmp\ptermR)
      \enskip.
      \end{equation*}
      Since
        $\height{\pterm[N_i']}<\height{\pterm[S_i]}\leq\height{\ptermQ}$,
      it holds that $\pterm[N_i']\pcmp\ptermR \prec\ptermP$
      and hence by the induction hypothesis there exists an
      \hsax{}-normal form $\pterm[N]$ such that
        $\pterm[N_i']\pcmp\ptermR\feqhs\pterm[N]$
      and
        $\height{\pterm[N_i']\pcmp\ptermR}\geq\height{\pterm[N]}$.
      Clearly,
        $\pref{\act}\pterm[N]$
      is an \hsax{}-normal form such that
        $\pterm[S_i]\lmerge\pterm[R]
           \feqhs\pref{\act}\pterm[N]$
      and
        $\height{\pterm[S_i]\lmerge\pterm[R]}\geq
           \height{\pref{\act}\pterm[N]}$.
    \item If $\pterm[S_i]=\pterm[S_i']\lmerge\pterm[N_i'']$ with
      $\pterm[S_i']=\varx$, $\pterm[S_i']=(\varx\cmerge\acta)$ or
      $\pterm[S_i']=(\varx\cmerge\vary)$, and $\pterm[N_i'']$ an
      \hsax{}-normal form, then by \lmlm{}
      \begin{equation*}
        \pterm[S_i]\lmerge\ptermR
          \feqhs\pterm[S_i']\lmerge(\pterm[N_i'']\pcmp\ptermR)
      \enskip.
      \end{equation*}
      Note that $\height{\pterm[S_i']}>0$, so
        $\height{\pterm[N_i'']}<\height{\pterm[S_i]}\leq\height{\ptermQ}$.
      It follows that $\pterm[N_i'']\pcmp\ptermR\prec\ptermP$, and
      hence by the induction hypothesis there exists an \hsax{}-normal
      form $\pterm[N]$ such that
        $\pterm[N_i'']\pcmp\ptermR\feqhs\pterm[N]$
      and
        $\height{\pterm[N_i'']\pcmp\ptermR}\geq\height{\pterm[N]}$.
      Clearly,
        $\pterm[S_i']\lmerge\pterm[N]$
      is an \hsax{}-normal form such that
        $\pterm[S_i]\lmerge\pterm[R]
           \feqhs\pterm[S_i']\lmerge\pterm[N]$
      and
        $\height{\pterm[S_i]\lmerge\pterm[R]}\geq
           \height{\pterm[S_i']\lmerge\pterm[N]}$.
    \end{enumerate}

  \item \label{case:hcmerge}
    If $\ptermP=\pterm[Q]\cmerge\pterm[R]$, then, since
    $\height{\pterm[Q]}\leq\height{\ptermP}$ and
    $\length{\pterm[Q]}<\length{\ptermP}$, it holds that
    $\pterm[Q]\prec\ptermP$, and, for similar reasons,
    $\pterm[R]\prec\ptermP$.
    Hence, by the induction hypothesis and Lemma~\ref{lem:simplify}
    there exist collections $\pterm[S_1],\dots,\pterm[S_m]$ and
    $\pterm[T_1],\dots,\pterm[T_n]$ of simple \hsax{}-normal forms
    such that
      $\ptermQ\feqhs\sum_{i=1}^m\pterm[S_i]$,
      $\ptermR\feqhs\sum_{j=1}^n\pterm[T_j]$,
      $\height{\ptermQ}\geq\height{\pterm[S_i]}$ for all $i=1,\dots,m$,
    and
      $\height{\ptermR}\geq\height{\pterm[T_j]}$ for all
      $j=1,\dots,n$.
    Note that if $m=0$, then
      $\ptermP\feqhs\nil\cmerge\ptermR\fequate\nil$ by \nilcm{},
    and if $n=0$, then
      $\ptermP\feqhs\ptermQ\cmerge\nil
              \feqhs\nil\cmerge\ptermQ
              \feqhs\nil$ by \cmcom{} and \nilcm{},
    and clearly $\height{\ptermP}\geq\height{\nil}$.
    Otherwise, by \accm{} and \cmcom{}
    \begin{equation*}
      \ptermP
        \feqhs\sum_{i=1}^m\sum_{j=1}^n(\pterm[S_i]\cmerge\pterm[T_j])
    \enskip,
    \end{equation*}
    and it remains to show, for all $i=1,\dots,m$ and $j=1,\dots,n$,
    that $\pterm[S_i]\cmerge\pterm[T_j]$ is provably equal to an
    appropriate \hsax{}-normal form. We distinguish cases according to
    the syntactic forms that $\pterm[S_i]$ and $\pterm[T_j]$ may take:
    \begin{enumerate}
    \item Suppose $\pterm[S_i]=\pref{\silent}\pterm[S_i']$; then
      $\pterm[S_i]\cmerge\pterm[T_j]\feqhs\nil$ by
      Lemma~\ref{lem:silentcom}, and clearly
        $\height{\pterm[S_i]\cmerge\pterm[T_j]}\geq 0$.
    \item Suppose $\pterm[T_j]=\pref{\silent}\pterm[T_j']$; then
      we apply \cmcom{} and proceed as in the previous case.
    \item Suppose $\pterm[S_i]=\pterm[S_i']\lmerge\pterm[S_i'']$
      with
        $\pterm[S_i']=\varx\cmerge\acta$
      or
        $\pterm[S_i']=\varx\cmerge\vary$;
      then by \cmlm{}, \cmass{}, \hsax{}, and \nillm{}
      \begin{equation*}
        \pterm[S_i]\cmerge\pterm[T_j]
          \fequate(\pterm[S_i']\cmerge\pterm[T_j])\lmerge\pterm[S_i'']
          \feqhs\nil\lmerge\pterm[S_i'']
          \fequate\nil
      \enskip,
      \end{equation*}
      and clearly
        $\height{\pterm[S_i]\cmerge\pterm[T_j]}\geq\height{\nil}$.
    \item Suppose $\pterm[T_j]=\pterm[T_j']\lmerge\pterm[T_j'']$
      with
        $\pterm[T_j']=\varx\cmerge\acta$
      or
        $\pterm[T_j']=\varx\cmerge\vary$; then
      $\pterm[S_i]\cmerge\pterm[T_j]
         \fequate\pterm[T_j]\cmerge\pterm[S_i]$ by \cmcom{}
      and we can proceed as in the previous case.
    \item Suppose $\pterm[S_i]=\pref{\acta}\pterm[S_i']$
          and $\pterm[T_j]=\pref{\actb}\pterm[T_j']$.

          If $\actb\not=\coacta$, then
            $\pterm[S_i]\cmerge\pterm[T_j]\fequate\nil$
              by \prefcmu{}
          and
            $\height{\pterm[S_i]\cmerge\pterm[T_j]}\geq\height{\nil}$.

          On the other hand, if $\actb=\coacta$, then
            $\pterm[S_i]\cmerge\pterm[T_j]
               \fequate\pref{\silent}(\pterm[S_i']\pcmp\pterm[T_j'])$
               by $\prefcmd{}$,
          and, since
            $\height{\pterm[S_i']}<\height{\pterm[S_i]}
                                  \leq\height{\pterm[Q]}$
          and
            $\height{\pterm[T_j']}<\height{\pterm[T_i]}
                                  \leq\height{\pterm[R]}$,
          it follows that
          $\pterm[S_i']\pcmp\pterm[T_j']\prec\ptermP$.
          So, by the induction hypothesis there exists an
          \hsax{}-normal form $\pterm[N]$ such that
            $\pterm[S_i']\pcmp\pterm[T_j']\feqhs\pterm[N]$
          and
            $\height{\pterm[S_i']\pcmp\pterm[T_j']}
               \geq
             \height{\pterm[N]}$.
          Then clearly $\pref{\silent}\pterm[N]$ is an \hsax{}-normal
          form such that
            $\pterm[S_i]\cmerge\pterm[T_j]
               \feqhs\pref{\silent}\pterm[N]$
          and
            $\height{\pterm[S_i]\cmerge\pterm[T_j]}
               \geq\height{\pref{\silent}\pterm[N]}$.
    \item Suppose $\pterm[S_i]=\pref{\acta}\pterm[S_i']$
          and $\pterm[T_j]=\varx\lmerge\pterm[T_j']$.
          Then
          \begin{alignat*}{2}
            \pref{\acta}\pterm[S_i']\cmerge(\varx\lmerge\pterm[T_j'])
          & \fequate
            \pref{\acta}(\nil\pcmp\pterm[S_i'])
              \cmerge(\varx\lmerge\pterm[T_j']) 
          &&\qquad\text{(by \pcmpidp{}, \pcmpcom{} in Lemma~\ref{lem:pcmpmonoid})} \\
          & \fequate
            (\acta\lmerge\pterm[S_i'])
              \cmerge(\varx\lmerge\pterm[T_j']) 
          &&\qquad\text{(by \preflm{})} \\
          & \fequate
            (\varx\cmerge\acta)\lmerge(\pterm[S_i']\pcmp\pterm[T_j'])
          &&\qquad\text{(by Lemma~\ref{lem:lmcmlm} and \cmcom{}).}
          \end{alignat*}
          Since
            $\height{\pterm[S_i']}<\height{\pterm[S_i]}
                                  \leq\height{\pterm[Q]}$
          and
            $\height{\pterm[T_j']}<\height{\pterm[T_i]}
                                  \leq\height{\pterm[R]}$,
          it follows that
            $\pterm[S_i']\pcmp\pterm[T_j']\prec\ptermP$,
          and hence by the induction hypothesis there exists an
          \hsax{}-normal form $\pterm[N]$ such that
            $\pterm[S_i']\pcmp\pterm[T_j']\feqhs\pterm[N]$
          and
            $\height{\pterm[S_i']\pcmp\pterm[T_j']}
               \geq
             \height{\pterm[N]}$.
          Then clearly $(\varx\cmerge\acta)\lmerge\pterm[N]$ is
          an \hsax{}-normal form such that
            $\pterm[S_i]\cmerge\pterm[T_j]
               \feqhs(\varx\cmerge\acta)\lmerge\pterm[N]$
          and
            $\height{\pterm[S_i]\cmerge\pterm[T_j]}
               \geq\height{(\varx\cmerge\acta)\lmerge\pterm[N]}$.
    \item If $\pterm[S_i]=\varx\lmerge\pterm[S_i']$
          and $\pterm[T_j]=\pref{\acta}\pterm[T_j']$,
          then the proof is analogous to the previous case.
    \item Suppose $\pterm[S_i]=\varx\lmerge\pterm[S_i']$
          and $\pterm[T_j]=\vary\lmerge\pterm[T_j']$.
          Then, by the derived equation \lmcmlm{} (see Lemma~\ref{lem:lmcmlm})
          \begin{equation*}
            \pterm[S_i]\cmerge\pterm[T_j]
              \fequate
                (\varx\cmerge\vary)\lmerge(\pterm[S_i']\pcmp\pterm[T_j'])
          \enskip.
          \end{equation*}
          Since
            $\height{\pterm[S_i']}<\height{\pterm[S_i]}
                                  \leq\height{\pterm[Q]}$
          and
            $\height{\pterm[T_j']}<\height{\pterm[T_i]}
                                  \leq\height{\pterm[R]}$,
          it follows that
            $\pterm[S_i']\pcmp\pterm[T_j']\prec\ptermP$,
          and hence by the induction hypothesis there exists an
          \hsax{}-normal form $\pterm[N]$ such that
            $\pterm[S_i']\pcmp\pterm[T_j']\feqhs\pterm[N]$
          and
            $\height{\pterm[S_i']\pcmp\pterm[T_j']}
               \geq
             \height{\pterm[N]}$.
          Then clearly $(\varx\cmerge\vary)\lmerge\pterm[N]$ is
          an \hsax{}-normal form such that
            $\pterm[S_i]\cmerge\pterm[T_j]
               \feqhs(\varx\cmerge\vary)\lmerge\pterm[N]$
          and
            $\height{\pterm[S_i]\cmerge\pterm[T_j]}
               \geq\height{(\varx\cmerge\vary)\lmerge\pterm[N]}$.
    \end{enumerate}

  \item
    If $\ptermP=\ptermQ\pcmp\ptermR$, then
      $\ptermP
         \fequate\ptermQ\lmerge\ptermR
                   \acmp
                 \ptermR\lmerge\ptermQ
                   \acmp
                 \ptermQ\cmerge\ptermR$.
    We can now proceed as in case~\ref{case:hlmerge} to show that for
    $\ptermQ\lmerge\ptermR$ and $\ptermR\lmerge\ptermQ$ there exist
    \hsax{}-normal forms $\pterm[N_1]$ and $\pterm[N_2]$,
    respectively, such that
      $\ptermQ\lmerge\ptermR\feqhs\pterm[N_1]$,
      $\ptermR\lmerge\ptermQ\feqhs\pterm[N_2]$,
      $\height{\ptermQ\lmerge\ptermR}\geq\height{\pterm[N_1]}$
    and
      $\height{\ptermR\lmerge\ptermQ}\geq\height{\pterm[N_2]}$.
    Furthermore, we can proceed as in case~\ref{case:hcmerge} to show
    that for $\ptermQ\cmerge\ptermR$ there exists an \hsax{}-normal
    form $\pterm[N_3]$ such that
      $\ptermQ\cmerge\ptermR\feqhs\pterm[N_3]$
    and
      $\height{\ptermQ\cmerge\ptermR}\geq\height{\pterm[N_3]}$.
    Then $\pterm[N_1]\acmp\pterm[N_2]\acmp\pterm[N_3]$ is an
    \hsax{}-normal form such that
      $\ptermP\feqhs\pterm[N_1]\acmp\pterm[N_2]\acmp\pterm[N_3]$
    and
      $\height{\ptermP}
         \geq\height{\pterm[N_1]\acmp\pterm[N_2]\acmp\pterm[N_3]}$.
  \qed
  \end{enumerate}
  \end{proof}

  We proceed to establish that for every two \hsax{}-normal forms
  $\pterm[N_1]$ and $\pterm[N_2]$ there exists a valuation
    $\valstar:\VAR\rightarrow\Proc$
  such that
  \begin{equation} \label{eq:hsgoal}
     \text{if}\
       \fevalstar{\pterm[N_1]}=\fevalstar{\pterm[N_2]},\ 
     \text{then}\ \pterm[N_1]\feqhs\pterm[N_2].
  \end{equation}
  The distinguishing valuations $\hvalstar$ will have a slightly more
  complicated definition than before, because of the more complicated
  notion of normal form.

  As in the previous section, the definition of $\hvalstar$ is
  parametrised with a natural number $\HSWidth$. Since $\cmerge$ may
  now occur in \hsax{}-normal forms, we also need to make sure that
  whatever process $\hvalstar$ assigns to variables has sufficient
  communication abilities. To achieve this, we also parametrise
  $\hvalstar$ with a finite subset
    $\HSAct=\{\act[a_1],\dots,\act[a_n]\}$
  of $\Act$ that is closed under the bijection $\bar{.}$ on
  $\Act$. (Note that every finite subset of $\Act$ has a finite
  superset with the aforementioned property.)  Based on $\HSWidth$ and
  $\HSAct$ we define the valuation $\hvalstar:\VAR\rightarrow\Proc$ by
  \begin{equation*}
    \hvalstar(\varx) =
      \pref{\act[a_1]}\brancher[(1\cdot\hpvarnum{\varx})]
        \acmp\cdots\acmp
      \pref{\act[a_n]}\brancher[(n\cdot\hpvarnum{\varx})]
  \enskip.
  \end{equation*}
  We shall prove that $\hvalstar$ satisfies Eqn.\ \eqref{eq:hsgoal}
  if the actions occurring in $\pterm[N_1]$ and $\pterm[N_2]$ are in
  $\HSAct\cup\{\silent\}$ and the widths of
  $\pterm[N_1]$ and $\pterm[N_2]$, defined below, do not exceed
  $\HSWidth$. We must also be careful to define the injection
  $\hpvarnum{\_}$ in such a way that the extra factors $1,\dots,n$ in
  the definition of $\hvalstar$ do not interfere with the numbers
  assigned to variables; we let $\hpvarnum{\_}$ denote an injection 
  \begin{equation*}
    \hpvarnum{\_}:
      \VAR\rightarrow
        \{m:\text{$m$ a prime number such that $m>n$ and $m>\HSWidth$}\}
  \end{equation*}
  that associates with every variable a prime number greater than the
  cardinality of $\HSAct$ and greater than $\HSWidth$.

  The definition of width also needs to take into account the
  cardinality of $\HSAct$ to maintain that the maximal branching
  degree in $\hevalstar{\ptermN}$ does not exceed $\hwidth{\ptermN}$.
  \begin{definition} \label{def:hwidth}
    We define the \emph{width} $\hwidth{\ptermN}$ of an \hsax{}-normal
    form $\ptermN$ as follows:
    \begin{enumerate}
    \renewcommand{\theenumi}{\roman{enumi}}
    \renewcommand{\labelenumi}{(\theenumi)}
    \item if $\ptermN=\nil$, then $\hwidth{\ptermN}=0$;
    \item if $\ptermN=\pterm[N_1]\acmp\pterm[N_2]$, then
            $\hwidth{\pterm[N]}=\hwidth{\pterm[N_1]}+\hwidth{\pterm[N_2]}$;
    \item if $\ptermN=\pref{\act}\pterm[N']$, then
      $\hwidth{\ptermN}=\max(\hwidth{\pterm[N']},1)$;
    \item if $\ptermN=\varx\lmerge\pterm[N']$, then
      $\hwidth{\ptermN}=\max(\hwidth{\pterm[N']},n)$;
    \item if $\ptermN=(\varx\cmerge\acta)\lmerge\pterm[N']$, then
      $\hwidth{\ptermN}=\max(\hwidth{\pterm[N']},1)$; and
    \item if $\ptermN=(\varx\cmerge\vary)\lmerge\pterm[N']$, then
      $\hwidth{\ptermN}=\max(\hwidth{\pterm[N']},n)$.
    \end{enumerate}
  \end{definition}

  \begin{lemma} \label{lem:bdegnf}
    For every \hsax{}-normal form $\ptermN$, the branching degree of
    $\hevalstar{\ptermN}$ is at most $\hwidth{\ptermN}$.
  \end{lemma}
  \begin{proof}
    Structural induction on $\ptermN$.
  \qed
  \end{proof}

  \begin{lemma} \label{lem:hrs}
    Let $\pterm[S]$ be a simple \hsax{}-normal form, let $\act\in\Acts$,
    and let $\procp$ be a process such that
      $\hevalstar{\pterm[S]}\step{\act}\procp$.
    Then the following statements hold:
    \begin{enumerate}
    \renewcommand{\theenumi}{\roman{enumi}}
    \renewcommand{\labelenumi}{(\theenumi)}
    \item \label{lem:hrs:item:pref}
      if $\ptermS=\pref{\act[\beta]}\pterm[N]$, then
      $\act=\act[\beta]$ and $\proc[p]=\hevalstar{\pterm[N]}$;
    \item \label{lem:hrs:item:varlmerge}
      if $\ptermS=\varx\lmerge\pterm[N]$, then
        $\act=\act[a_i]$
      and
        $\proc[p]=\brancher[i\cdot\hpvarnum{\varx}]
                    \pcmp\hevalstar{\ptermN}$
      for some $i\in\{1,\dots,n\}$;
    \item \label{lem:hrs:item:varactcmerge}
      if $\ptermS=(\varx\cmerge\acta)\lmerge\pterm[N]$, then
        $\act=\silent$
      and
        $\proc[p]=\brancher[i\cdot\hpvarnum{\varx}]
                    \pcmp\hevalstar{\ptermN}$
        for the unique $i\in\{1,\dots,n\}$ such that
        $\coacta=\act[a_i]$; and
    \item \label{lem:hrs:item:varvarcmerge}
      if $\ptermS=(\varx\cmerge\vary)\lmerge\pterm[N]$, then
        $\act=\silent$
      and
        $\proc[p]=\brancher[i\cdot\hpvarnum{\varx}]
                    \pcmp
                  \brancher[j\cdot\hpvarnum{\vary}]
                    \pcmp\hevalstar{\ptermN}$
      for some $i,j\in\{1,\dots,n\}$ such that $\coact[a_i]=\act[a_j]$.
    \end{enumerate}
  \end{lemma}

  As in the previous section, we distinguish \hsax{}-normal forms by
  classifying their residuals according to the number of parallel
  components with a branching degree that exceeds $\HSWidth$.
  Again, we say
  that a process $\procp$ is of \emph{type $n$} ($n\geq 0$) if its
  unique parallel decomposition contains precisely $n$ parallel prime
  components with a branching degree $>\HSWidth$.
  \begin{corollary} \label{cor:hrt}
    Let $\ptermS$ be a simple \hsax{}-normal form such that
    $\hwidth{\pterm[S]}\leq\HSWidth$ and such that the actions occurring in
    $\ptermS$ are included in $\HSAct\cup\{\silent\}$.
    \begin{enumerate}
    \renewcommand{\theenumi}{\roman{enumi}}
    \renewcommand{\labelenumi}{(\theenumi)}
    \item \label{cor:hrt:item:pref}
      If $\ptermS=\pref{\act}\ptermN$, then the unique residual of
      $\hevalstar{\ptermS}$ is of type $0$.
    \item \label{cor:hrt:item:varlmerge}
      If $\ptermS=\varx\lmerge\ptermN$, then all residuals of
      $\hevalstar{\ptermS}$ are of type $1$.
    \item \label{cor:hrt:item:varactcmerge}
      If $\ptermS=(\varx\cmerge\acta)\lmerge\ptermN$, then
      the unique residual of $\hevalstar{\ptermS}$ is of type $1$.
    \item \label{cor:hrt:item:varvarcmerge}
      If $\ptermS=(\varx\cmerge\vary)\lmerge\ptermN$, then all
      residuals of $\hevalstar{\ptermS}$ are of type $2$.
    \end{enumerate}
  \end{corollary}
  \begin{proof}
    On the one hand, by Lemma~\ref{lem:bdegnf}, in each case
    $\hevalstar{\pterm[N]}$ has a branching degree of at most
    $\hwidth{\ptermN}\leq\fwidth{\ptermS}\leq\HSWidth$, and hence, by
    Lemma~\ref{lem:bdegmergea}, its unique parallel decomposition
    cannot contain parallel prime components with a branching degree
    that exceeds $\HSWidth$. On the other hand, by
    Lemmas~\ref{lem:brancher}(\ref{lem:brancher:item:prime}) and
    \ref{lem:brancher}(\ref{lem:brancher:item:bdeg}), the processes
      $\brancher[i\cdot\hpvarnum{\varx}]$
    and
      $\brancher[j\cdot\hpvarnum{\vary}]$
    are parallel prime and have a branching degree that exceeds
    $\HSWidth$.
    Further note that, since the assumption on \CCS{} communication
    functions that $\co\acta\not=\acta$ implies that $i\neq j$, the
    processes
      $\brancher[i\cdot\hpvarnum{\varx}]$
    and
      $\brancher[j\cdot\hpvarnum{\vary}]$
    are distinct.
    Using these observations it is straightforward to establish the
    corollary as a consequence of
    Lemma~\ref{lem:hrs}.
  \qed
  \end{proof}

  \begin{theorem} \label{theo:hnfdistinction}
    For every two \hsax{}-normal forms $\pterm[N_1]$, $\pterm[N_2]$ such
    that $\hwidth{\pterm[N_1]},\hwidth{\pterm[N_2]}\leq W$ and such that
    the actions occurring in $\pterm[N_1]$ and $\pterm[N_2]$ are included
    in $\HSAct\cup\{\silent\}$ it holds that
      $\hevalstar{\pterm[N_1]}=\hevalstar{\pterm[N_2]}$
    only if
      $\pterm[N_1]\fequate\pterm[N_2]$ modulo \accom{}--\acnil{},
      \cmcom{}.
  \end{theorem}
  \begin{proof}
    By Lemma~\ref{lem:simplify} we may assume that $\pterm[N_1]$ and
    $\pterm[N_2]$ are summations of collections of simple \hsax{}-normal
    forms.
    We assume $\hevalstar{\pterm[N_1]}=\hevalstar{\pterm[N_2]}$ and
    prove that then $\pterm[N_1]\fequate\pterm[N_2]$ modulo
    \accom{}--\acnil{}, \cmcom{}, by induction on the sum of the
    heights of $\pterm[N_1]$ and $\pterm[N_2]$.

    We first prove that for every syntactic summand $\pterm[S_1]$ of
    $\pterm[N_1]$ there is a syntactic summand $\pterm[S_2]$ of
    $\pterm[N_2]$ such that $\pterm[S_1]\fequate\pterm[S_2]$ modulo 
    \accom{}--\acnil{}, \cmcom{}.
    To this end, let $\pterm[S_1]$ be an arbitrary syntactic summand
    of $\pterm[N_1]$; we distinguish cases according to the syntactic
    form of $\pterm[S_1]$.
    \begin{enumerate}
    \item Suppose $\pterm[S_1]=\pref{\act}\pterm[N_1']$;
       then
         $\hevalstar{\pterm[S_1]}\step{\act}\hevalstar{\pterm[N_1']}$.
       Hence, since $\hevalstar{\pterm[N_1]}=\hevalstar{\pterm[N_2]}$,
       there exists a syntactic summand $\pterm[S_2]$ of $\pterm[N_2]$
       such that
         $\hevalstar{\pterm[S_2]}\step{\act}\hevalstar{\pterm[N_1']}$.
       By Lemma~\ref{lem:bdegnf} the branching degree of
       $\hevalstar{\pterm[N_1']}$ does not exceed $\HSWidth$, so
       $\hevalstar{\pterm[S_2]}$ has a residual of type $0$, and
       therefore, by Corollary~\ref{cor:hrt}, there exist
       $\act[\beta]\in\Acts$ and an \hsax{}-normal form $\pterm[N_2']$
       such that
         $\pterm[S_2]=\pref{\act[\beta]}\pterm[N_2']$.
       Moreover, since
         $\hevalstar{\pterm[S_2]}\step{\act}\hevalstar{\pterm[N_1']}$
       it follows by
       Lemma~\ref{lem:hrs}(\ref{lem:hrs:item:pref})
       that $\act=\act[\beta]$ and
       $\hevalstar{\pterm[N_1']}=\hevalstar{\pterm[N_2']}$.
       Hence, by the induction hypothesis, we conclude that
         $\pterm[N_1']\fequate\pterm[N_2']$ modulo \accom{}--\acnil{},
         \cmcom{}.
       So $\pterm[S_1]=\pref{\act}\pterm[N_1']
             \fequate
           \pref{\act[\beta]}\pterm[N_2']=\pterm[S_2]$.
    \item Suppose $\pterm[S_1]=\varx\lmerge\pterm[N_1']$;
      then
        $\hevalstar{\pterm[S_1]}
           \step{\act[a_1]}
             \brancher[\hpvarnum{\varx}]\pcmp\hevalstar{\pterm[N_1']}$.
      Hence, since $\hevalstar{\pterm[N_1]}=\hevalstar{\pterm[N_2]}$,
      there exists a summand $\pterm[S_2]$ of $\pterm[N_2]$ such that
        $\hevalstar{\pterm[S_2]}
           \step{\act[a_1]}
             \brancher[\hpvarnum{\varx}]\pcmp\hevalstar{\pterm[N_1']}$.
      Since $\pterm[S_2]$ has a residual of type 1, by
      Corollary~\ref{cor:hrt}(\ref{cor:hrt:item:pref},
      \ref{cor:hrt:item:varvarcmerge}) it is not of the form
      $\pref{\act}\pterm[N_2']$ for some $\act\in\Acts$ and
      \hsax{}-normal form $\pterm[N_2']$, or of the form
      $(\vary\cmerge\varz)\lmerge\pterm[N_2']$ for some
      $\vary,\varz\in\VAR$ and \hsax{}-normal form $\pterm[N_2']$.
      Moreover, $\pterm[S_2]$ cannot be of the form
        $(\vary\cmerge\acta)\lmerge\pterm[N_2']$ for some
        $\vary\in\VAR$ and $\acta\in\Act$,
      for then by Lemma~\ref{lem:hrs}(\ref{lem:hrs:item:varactcmerge})
        $\hevalstar{\pterm[S_2]}\step{\act}\proc[p]$
      would imply $\act=\silent\not=\act[a_1]$.
      So, there exists a variable $\vary$ and an \hsax{}-normal form
      $\pterm[N_2']$ such that
        $\pterm[S_2]=\vary\lmerge\pterm[N_2']$.
      Now, since
        $\hevalstar{S_2}
           \step{\act[a_1]}
             \brancher[\hpvarnum{\varx}]\pcmp\hevalstar{\pterm[N_1']}$,
      it follows by Lemma~\ref{lem:hrs}(\ref{lem:hrs:item:varlmerge})
      that
      \begin{equation} \label{eq:paralleleq1}
         \brancher[\hpvarnum{\varx}]\pcmp\hevalstar{\pterm[N_1']}
       = \brancher[\hpvarnum{\vary}]\pcmp\hevalstar{\pterm[N_2']}
      \enskip.
      \end{equation}
      Since $\hevalstar{\pterm[N_1']}$ and $\hevalstar{\pterm[N_2']}$
      are of type $0$, we conclude that the unique decomposition of
      $\hevalstar{\pterm[N_1']}$ does not contain
      $\brancher[\hpvarnum{\vary}]$ and the unique decomposition of
      $\hevalstar{\pterm[N_2']}$ does not contain
      $\brancher[\hpvarnum{\varx}]$.
      Hence, from \eqref{eq:paralleleq1} it follows that
        $\brancher[\hpvarnum{\varx}]=\brancher[\hpvarnum{\vary}]$
      and
        $\hevalstar{\pterm[N_1']}=\hevalstar{\pterm[N_2']}$.
      From the former we conclude by the injectivity of $\hpvarnum{.}$
      that $\varx=\vary$, and from the latter
      we conclude by the induction hypothesis that
        $\pterm[N_1']\fequate\pterm[N_2']$ modulo \accom{}--\acnil{},
        \cmcom{}.
      So $\pterm[S_1]=\varx\lmerge\pterm[N_1']
            \fequate
          \vary\lmerge\pterm[N_2']=\pterm[S_2]$.
    \item \label{theo:hnfdistinction:case:varactlmerge}
      Suppose
      $\pterm[S_1]=(\varx\cmerge\acta)\lmerge\pterm[N_1']$,
      and let $i$ be such that $\coacta=\act[a_i]$.
      Then
         $\hevalstar{\pterm[S_1]}
            \step{\silent}
              \brancher[i\cdot\hpvarnum{\varx}]
                \pcmp\hevalstar{\pterm[N_1']}$.
      Hence, since $\hevalstar{\pterm[N_1]}=\hevalstar{\pterm[N_2]}$,
      there exists a summand $\pterm[S_2]$ of $\pterm[N_2]$ such that
      \begin{equation*}
        \hevalstar{\pterm[S_2]}
          \step{\silent}
            \brancher[i\cdot\hpvarnum{\varx}]
              \pcmp\hevalstar{\pterm[N_1']}
      \enskip.
      \end{equation*}
      Since $\pterm[S_2]$ has a residual of type 1, by
      Corollary~\ref{cor:hrt}(\ref{cor:hrt:item:pref},\ref{cor:hrt:item:varvarcmerge}) it is not of the form
      $\pref{\act}\pterm[N_2']$ for some $\act\in\Acts$ and
      \hsax{}-normal form $\pterm[N_2']$, or of the form
      $(\vary\cmerge\varz)\lmerge\pterm[N_2']$ for some
      $\vary,\varz\in\VAR$ and \hsax{}-normal form $\pterm[N_2']$.
      Moreover, $\pterm[S_2]$ cannot be of the form
        $\vary\lmerge\pterm[N_2']$ for some $\vary\in\VAR$,
      for then by Lemma~\ref{lem:hrs}(\ref{lem:hrs:item:varlmerge})
        $\hevalstar{\pterm[S_2]}\step{\act}\procp$
      would imply $\act=\act[a_k]\not=\silent$ for some
      $k\in\{1,\dots,n\}$.
      So, there exist a variable $\vary$, action $\actb\in\HSAct$
      and an \hsax{}-normal form $\pterm[N_2']$ such that
        $\pterm[S_2]=(\vary\cmerge\actb)\lmerge\pterm[N_2']$.
      Now, since
        $\hevalstar{S_2}
           \step{\silent}
             \brancher[i\cdot\hpvarnum{\varx}]\pcmp\hevalstar{\pterm[N_1']}$,
      it follows by
      Lemma~\ref{lem:hrs}(\ref{lem:hrs:item:varactcmerge}) that
      \begin{equation} \label{eq:paralleleq2}
         \brancher[i\cdot\hpvarnum{\varx}]\pcmp\hevalstar{\pterm[N_1']}
       =
         \brancher[j\cdot\hpvarnum{\vary}]\pcmp\hevalstar{\pterm[N_2']}
      \enskip,
      \end{equation}
      with $j\in\{1,\dots,n\}$ such that $\coactb=\act[a_j]$.
      By
        Lemma~\ref{lem:brancher}(\ref{lem:brancher:item:prime},\ref{lem:brancher:item:bdeg})
      the processes
        $\brancher[i\cdot\hpvarnum{\varx}]$
      and
        $\brancher[j\cdot\hpvarnum{\vary}]$
      are parallel prime and have branching degrees that, since
        $\hpvarnum{\varx}>\HSWidth$ and $\hpvarnum{\vary}>\HSWidth$,
      exceed $\HSWidth$.
      Therefore, since $\hevalstar{\pterm[N_1']}$ and
      $\hevalstar{\pterm[N_2']}$ are of type $0$, it follows that the
      unique decomposition of $\hevalstar{\pterm[N_1']}$ does not contain
      $\brancher[j\cdot\hpvarnum{\vary}]$ and the unique decomposition of
      $\hevalstar{\pterm[N_2']}$ does not contain
      $\brancher[i\cdot\hpvarnum{\varx}]$.
      Hence, by \eqref{eq:paralleleq2} we have that
        $\brancher[i\cdot\hpvarnum{\varx}]=\brancher[j\cdot\hpvarnum{\vary}]$
      and
        $\hevalstar{\pterm[N_1']}=\hevalstar{\pterm[N_2']}$.
      From
        $\brancher[i\cdot\hpvarnum{\varx}]
           =\brancher[j\cdot\hpvarnum{\vary}]$,
      by Lemma~\ref{lem:brancher}(\ref{lem:brancher:item:distinct})
      we infer that
        $i\cdot\hpvarnum{\varx}=j\cdot\hpvarnum{\vary}$.
      Since $\hpvarnum{\varx}$ and $\hpvarnum{\vary}$ are prime
      numbers greater than $i$ and $j$, it follows that $i=j$, whence
      $\acta=\actb$, and $\hpvarnum{\varx}=\hpvarnum{\vary}$, whence
      $\varx=\vary$ by the injectivity of $\hpvarnum{.}$.
      From $\hevalstar{\pterm[N_1']}=\hevalstar{\pterm[N_2']}$ we
      conclude by the induction hypothesis that 
        $\pterm[N_1']\fequate\pterm[N_2']$ modulo \accom{}--\acnil{},
        \cmcom{}.
      So $\pterm[S_1]=(\varx\cmerge\acta)\lmerge\pterm[N_1']
            \fequate
          (\vary\cmerge\actb)\lmerge\pterm[N_2']=\pterm[S_2]$.
    \item Suppose
      $\pterm[S_1]=(\varx\cmerge\vary)\lmerge\pterm[N_1']$.
      Then
         $\hevalstar{\pterm[S_1]}
            \step{\silent}
              \brancher[i\cdot\hpvarnum{\varx}]
                \pcmp
              \brancher[j\cdot\hpvarnum{\vary}]
                \pcmp
              \hevalstar{\pterm[N_1']}$
      with $i,j\in\{1,\dots,n\}$ such that $\coact[a_i]=\act[a_j]$.
      Hence, since $\hevalstar{\pterm[N_1]}=\hevalstar{\pterm[N_2]}$,
      there exists a summand $\pterm[S_2]$ of $\pterm[N_2]$ such that
      \begin{equation*}
        \hevalstar{\pterm[S_2]}
          \step{\silent}
              \brancher[i\cdot\hpvarnum{\varx}]
                \pcmp
              \brancher[j\cdot\hpvarnum{\vary}]
                \pcmp
              \hevalstar{\pterm[N_1']}
      \enskip.
      \end{equation*}
      Since $\pterm[S_2]$ has a residual of type 2, by
      Corollary~\ref{cor:hrt} there exist
      $\var[x'],\var[y']\in\VAR$ and an \hsax{}-normal form
      $\pterm[N_2']$ such that
        $\pterm[S_2]=(\var[x']\cmerge\var[y'])\lmerge\pterm[N_2']$.
      Now, since
        $\hevalstar{S_2}
           \step{\silent}
              \brancher[i\cdot\hpvarnum{\varx}]
                \pcmp
              \brancher[j\cdot\hpvarnum{\vary}]
                \pcmp
              \hevalstar{\pterm[N_1']}$
      it follows by
      Lemma~\ref{lem:hrs}(\ref{lem:hrs:item:varvarcmerge}) that for
      some $k,l\in\{1,\dots,n\}$ such that $\coact[a_k]=\act[a_l]$
      \begin{equation} \label{eq:paralleleq3}
        \brancher[i\cdot\hpvarnum{\varx}]
          \pcmp
        \brancher[j\cdot\hpvarnum{\vary}]
          \pcmp
        \hevalstar{\pterm[N_1']}
       =
        \brancher[k\cdot\hpvarnum{\var[x']}]
          \pcmp
        \brancher[l\cdot\hpvarnum{\var[y']}]
          \pcmp
        \hevalstar{\pterm[N_2']}
      \enskip.
      \end{equation}
      By Lemma~\ref{lem:brancher}(\ref{lem:brancher:item:prime},\ref{lem:brancher:item:bdeg}) the processes
        $\brancher[i\cdot\hpvarnum{\varx}]$,
        $\brancher[j\cdot\hpvarnum{\vary}]$,
        $\brancher[k\cdot\hpvarnum{\var[x']}]$
      and
        $\brancher[l\cdot\hpvarnum{\var[y']}]$
      are parallel prime and have branching degrees that exceed
      $\HSWidth$.
      Therefore, since $\hevalstar{\pterm[N_1']}$ and
      $\hevalstar{\pterm[N_2']}$ are of type $0$, it follows that the
      unique decomposition of $\hevalstar{\pterm[N_1']}$ does not
      contain
        $\brancher[k\cdot\hpvarnum{\var[x']}]$
      and
        $\brancher[l\cdot\hpvarnum{\var[y']}]$,
      and the unique decomposition of
      $\hevalstar{\pterm[N_2']}$ does not contain
        $\brancher[i\cdot\hpvarnum{\varx}]$
      and
        $\brancher[j\cdot\hpvarnum{\vary}]$.
      Hence, from \eqref{eq:paralleleq3} we infer that
        $\hevalstar{\pterm[N_1']}=\hevalstar{\pterm[N_2']}$
      and either
        $\brancher[i\cdot\hpvarnum{\varx}]
           =\brancher[k\cdot\hpvarnum{\var[x']}]$
      and
        $\brancher[j\cdot\hpvarnum{\vary}]
           =\brancher[l\cdot\hpvarnum{\var[y']}]$,
      or
        $\brancher[i\cdot\hpvarnum{\varx}]
           =\brancher[l\cdot\hpvarnum{\var[y']}]$
      and
        $\brancher[j\cdot\hpvarnum{\vary}]
           =\brancher[k\cdot\hpvarnum{\var[x']}]$.

      From the former we conclude by the induction hypothesis that 
        $\pterm[N_1']\fequate\pterm[N_2']$ modulo \accom{}--\acnil{},
        \cmcom{};
      from the latter it follows reasoning as in case~\ref{theo:hnfdistinction:case:varactlmerge} that either $\varx=\var[x']$ and
      $\vary=\var[y']$, or $\varx=\var[y']$ and $\vary=\var[x']$.
      In both cases,
         $\pterm[S_1]=(\varx\cmerge\vary)\lmerge\pterm[N_1']
            \fequate
          (\var[x']\cmerge\var[y'])\lmerge\pterm[N_2']=\pterm[S_2]$.
    \end{enumerate}
    We have established that every syntactic summand of $\pterm[N_1]$
    is provably equal to a syntactic summand of
    $\pterm[N_2]$. Similarly, it follows that every syntactic summand
    of $\pterm[N_2]$ is provably equal to a syntactic summand of 
    $\pterm[N_2]$. Hence, modulo \accom{}--\acnil{}, \cmcom{}
      $\pterm[N_1]
         \fequate\pterm[N_1]\acmp\pterm[N_2]\fequate
           \pterm[N_2]$,
    and the proof of the theorem is complete.
  \qed
  \end{proof}

  \begin{corollary} \label{cor:hscomplete}
    For all process terms $\ptermP$ and $\ptermQ$,
      $\ptermP\feqhs\ptermQ$
    if, and only if,
      $\ptermP\bisimilarccs\ptermQ$,
    and hence the axioms generated by the schemata in
    Table~\ref{tab:axccs} together with the axiom \hsax{} consitute an
    equational base for $\ProcCCS{}$.
  \end{corollary}
  \begin{proof}
    The implication from left to right is
    Proposition~\ref{prop:ccssound}.
    To prove the implication from right to left, suppose
    $\ptermP\bisimilarccs\ptermQ$. Then, by Lemma~\ref{lem:hnf} there
    exist \hsax{}-normal forms $\pterm[N_1]$ and $\pterm[N_2]$ such
    that $\ptermP\feqhs\pterm[N_1]$ and $\ptermQ\feqhs\pterm[N_2]$;
    from $\ptermP\bisimilarccs\ptermQ$ we conclude by
    Proposition~\ref{prop:ccssound} that
      $\pterm[N_1]\bisimilarccs\pterm[N_2]$.
    Now choose $\HSWidth$ large enough such that
      $\hwidth{\pterm[N_1]},\hwidth{\pterm[N_2]}\leq W$, and pick a
    finite set $\HSAct$ that is closed under $\bar{.}$ and
      includes all of the actions occurring in $\pterm[N_1]$ and
      $\pterm[N_2]$.
    From $\pterm[N_1]\bisimilarccs\pterm[N_2]$ it follows that
    $\hevalstar{\pterm[N_1]}=\hevalstar{\pterm[N_2]}$, and hence, by
    Theorem~\ref{theo:hnfdistinction} $\pterm[N_1]\fequate\pterm[N_2]$.
    We can therefore conclude
      $\pterm[P]
         \feqhs
       \pterm[N_1]
         \fequate
       \pterm[N_2]
         \feqhs
       \pterm[Q]$.
  \qed
  \end{proof}

  \begin{corollary}
    The equational theory of \ProcCCS{} is decidable.
  \end{corollary}
  \begin{proof}
    From the proof of Lemma~\ref{lem:hnf} it is easy to see that
    there exists an effective procedure that associates with every
    process term a provably equivalent \hsax{}-normal.
    Furthermore, from Definition~\ref{def:hwidth} it is clear that,
    given a set \HSAct{}, every \hsax{}-normal form has an effectively
    computable width.
    We now sketch an effective procedure that decides whether a
    process equation
      $\ptermP\fequate\ptermQ$
    is valid:
    \begin{enumerate}
    \item Compute \hsax{}-normal forms $\pterm[N_1]$ and $\pterm[N_2]$
      such that $\ptermP\feqhs\pterm[N_1]$ and $\ptermQ\feqhs\pterm[N_2]$.
    \item Determine the least set
        $\HSAct{}=\{\act[a_1],\dots,\act[a_n]\}$
      of actions that is closed under $\bar{.}$ and contains the
      actions in $\Act$ occurring in $\pterm[N_1]$ and $\pterm[N_2]$.
    \item Compute $\hwidth{\pterm[N_1]}$ and $\hwidth{\pterm[N_2]}$ given
      \HSAct{}, and define $\FMWidth$ as their maximum.
    \item Determine the (finite) set $\VAR[V']$ of variables
      occurring in $\pterm[N_1]$ and $\pterm[N_2]$; define an
      injection
      \begin{equation*}
        \hpvarnum{.}:\VAR[V']\rightarrow
          \{m\in\N:\text{$m$ a prime number such that $m>n$ and $m>\FMWidth$}\}
      \enskip,
      \end{equation*}
      and a substitution
        $\hvalstar:\VAR[V']\rightarrow\CPTERMS$
      that assigns to a variable $\varx$ in $\VAR[V']$ the closed
      process term
      \begin{equation*}
        \pref{\act[a_1]}\brancher[1\cdot\hpvarnum{\varx}]
      \acmp\cdots\acmp
        \pref{\act[a_n]}\brancher[n\cdot\hpvarnum{\varx}]
      \enskip.
      \end{equation*}
      (Again, we interpret Eqn.~\eqref{eq:brancherdef} as
      defining a sequence of closed process terms instead of a
      sequence of processes.) 
    \item Let $\pterm[N_1^{\hvalstar}]$ and $\pterm[N_2^{\hvalstar}]$
      be the results from applying $\hvalstar$ to $\pterm[N_1]$ and
      $\pterm[N_2]$, respectively.
    \item Determine if the closed process terms
      $\pterm[N_1^{\hvalstar}]$ and $\pterm[N_2^{\hvalstar}]$ are
      bisimilar;
      if they are, then the process equation
        $\ptermP\fequate\ptermQ$
      is valid in \ProcCCS{}, and otherwise it is not.
\qed
    \end{enumerate}
  \end{proof}

  \section{Concluding remarks}

  We have discussed the equational theories of two process algebras
  arising from the fragment of \CCS{} without recursion, restriction
  and relabelling. Moller has proved in \cite{Mol90} that these
  equational theories are not finitely based. We have shown that if 
  the set of actions is finite and the auxiliary operators left merge
  and communication merge from Bergstra and Klop \cite{BK84} are
  added, then finite equational bases can be obtained. They
  consist of (adaptations of) axioms appearing already in
  \cite{BK84,BT85,HM85}.

  Denote by $\EqBase$ the set of the axioms generated by the
  schemata in Table~\ref{tab:axccs} on p.~\pageref{tab:axccs} together
  with the axiom
    $\varx\cmerge(\vary\cmerge\varz)\fequate\nil$,
  which expresses the communication mechanism conforms to the
  \emph{handshaking paradigm}. Our main result
  (Corollary~\ref{cor:hscomplete}) establishes that $\EqBase$ is an
  equational base for the algebra $\ProcCCS$. Note that an equational
  base for an algebra is an equational base for every extension of
  that algebra in which the axioms hold.%
    \footnote{The algebra $\mathbf{B}$ is an extension of the algebra
      $\mathbf{A}$ if there exists an embedding from $\mathbf{A}$ into
      $\mathbf{B}$.}
  So, as a consequence of our result, $\EqBase$ is in fact an
  equational base, e.g., for every algebra of process graphs
  modulo bisimulation endowed with a distinguished element $\nil$ and
  operations $\pref{\act}$ ($\act\in\Acts$), $\acmp$, $\pcmp$,
  $\lmerge$ and $\cmerge$ according to their standard
  interpretations. In particular it is clear from the preceding
  remarks that, although the algebra $\ProcCCS$ contains only finite
  processes, this is not essential for our result.

  As a special case of Corollary~\ref{cor:hscomplete}, the axiom
  system $\EqBase$ is \emph{ground-complete} with respect to
  bisimilarity (i.e., $\feqhs$ coincides with $\bisimilarccs$ on the
  set of \emph{closed} terms $\CPTERMS$). Consequently, the algebra
  $\ProcCCS$ is isomorphic with the \emph{initial algebra} associated
  with $\EqBase$, i.e., the quotient of the set of closed terms modulo
  $\feqhs$. It also follows from our main result that the axiom system
  is \emph{$\omega$-complete}. For suppose that every closed instance
  of the equation $\ptermP\fequate\ptermQ$ is derivable; then the
  equation itself is valid in the initial algebra. By
  ground-completeness, it follows that $\ptermP\fequate\ptermQ$ is
  valid in $\ProcCCS$, and hence, by Corollary~\ref{cor:hscomplete},
  it is derivable from $\EqBase$.

  As a stepping stone towards our main result, we first considered the
  process algebra $\ProcFM$ with a trivial communication
  mechanism. An equational base for it is obtained if the axiom
    $\varx\cmerge\vary\fequate\nil$
  is added to the axioms generated by the schemata in
  Table~\ref{tab:axccs} on p.~\pageref{tab:axccs}
  (Corollary~\ref{cor:fmcomplete}). The auxiliary operator $\cmerge$
  is then actually superfluous. For we can replace $\pcmplmcm{}$ by
    $\varx\pcmp\vary\fequate\varx\lmerge\vary\acmp\vary\lmerge\varx$,
  and, moreover, transform every equational proof into a proof in which
  $\cmerge$ does not occur by replacing every occurrence of a
  subexpression $\ptermP\cmerge\ptermQ$ by $\nil$. It follows that the
  axiomatisation consisting of \accom{}--\acnil{}, \nillm{}--\lmnil{},
  and the simplified axiom \pcmplmcm{} is $\omega$-complete. Thus, we
  generalise the result of Moller \cite{Mol89}, who establishes
  $\omega$-completeness of the axiomatisation under the condition
  that the set of actions is infinite; according to our result the
  condition can be omitted.

\end{document}